\documentclass[12pt]{article}
\usepackage{epsfig}
\usepackage{graphicx}
\usepackage{amsmath}
\usepackage{amsfonts}
\usepackage{dsfont}
\usepackage{eufrak}

\newcommand{\be}{\begin{equation}}
\newcommand{\ee}{\end{equation}}
\newcommand{\bea}{\begin{eqnarray}}
\newcommand{\eea}{\end{eqnarray}}

\newcommand{\mbf}[1]{\mbox{\boldmath $#1$}}

\newcommand{\bq}{\mbf{q}}

\newcommand{\la}{\lambda}

\newcommand{\brho}{\mbf{\rho}}

\begin{document}
DESY 05-103                           \hfill ISSN 0418-9833 \\

\begin{center}
{\Large \vspace*{0.5cm} \textbf{Deformed Spectral Representation of the BFKL
Kernel and the Bootstrap for Gluon Reggeization}\\[0pt]
} \vspace{0.5cm}

\renewcommand{\thefootnote}{\fnsymbol{footnote}} \renewcommand{%
\thefootnote}{\arabic{footnote}} \setcounter{footnote}{0}

{\large {J. Bartels$^{(a)}$,
%\footnote{\noindent email: krisztian.peters@desy.de}
\renewcommand{\thefootnote}{\fnsymbol{footnote}} \renewcommand{%
\thefootnote}{\arabic{footnote}}\hspace{-0.3cm} L.N. Lipatov$^{(b)}$, M.
Salvadore$^{(c)}$  and G.P. Vacca$^{(c)}$} }\\[0pt]
\vspace{0.3cm} \textit{$^{(a)}$II.\ Institut f\"ur Theoretische Physik,
Universit\"at Hamburg,\\[0pt]
Luruper Chaussee 149, D-22761 Hamburg, Germany} \\[0pt]
\vspace{0.2cm} \textit{$^{(b)}$ Petersburg Nuclear Physics Institute\\[0pt]
Gatchina, 188 300 St.Petersburg, Russia}\\[0pt]
\vspace{0.2cm} \textit{$^{(c)}$ Dipartimento di Fisica - Universit\`a di
Bologna and INFN - Sezione di Bologna,\\[0pt]
via Irnerio 46, 40126 Bologna, Italy}
\end{center}

\begin{abstract}
\noindent
We investigate the space of functions in which the BFKL kernel
acts. For the amplitudes which describe the scattering of
colorless projectiles it is convenient to define, in transverse
coordinates, the M\"obius space in which the solutions to the BFKL
equation vanish as the coordinates of the two reggeized gluons
coincide. However, in order to fulfill the bootstrap relation
for the BFKL kernel it is necessary to modify the space of
functions. We define and investigate a new space of functions
and show explicitly that the bootstrap relation is valid for the
corresponding spectral form of the kernel.
We calculate the generators of the resulting deformed representation
of the $\mathfrak{sl}(2,\mathbb{C})$ algebra.
\end{abstract}

%%%%%%%%%%%%%%%%%%%%%%%%%%%%%%%%%%%%%%%%%%%%%%%%%%%%%%%%%%%%%%%%%%%

%%%%%%%%%%%%%%%%%%%%%%%%%%%%%%%%%%%%%%%%%%%%%%%%%%%%%%%%%%%%%%%%%%%

\section{Introduction}

The leading order BFKL kernel~\cite{BFKL}, derived from Feynman diagrams in
momentum space, has been investigated in much detail. As a function of the
transverse momenta, it is a meromorphic function. In the color singlet
exchange channel, it describes the Pomeron contribution in pQCD,
consisting of ladder diagrams with reggeized gluons. This BFKL
Pomeron couples to the impact factors of colourless particles
which, because of gauge invariance, vanish as one of the two
reggeized gluons carries a zero transverse momentum. This
property allows us to modify the space of functions to which the
Pomeron wave function belongs.  On the other hand, in the color
octet exchange channel the BFKL equation has the bootstrap
solution, which represents a fundamental consistency property
derived from the s-channel unitarity. A manifestation of this
bootstrap property also takes place for the color singlet state
of two gluons and inside the coupling of three or more gluons to
colorless particle impact factors. In particular, it plays an
important role in the Odderon solution~\cite{BLV} which appears
as a bound state of three reggeized gluons.

An important virtue of the color singlet kernel is its
invariance under the M\"{o}bius transformations in the space of
transverse coordinates. Exploiting the gauge invariance of the
colorless impact factor to which the BFKL Pomeron couples, the
M\"{o}bius symmetry allows to search  solutions of the BFKL
eigenvalues equation in the space of M\"obius functions which
vanish as the coordinates of the two reggeized gluons coincide,
and to define a spectral representation of the BFKL kernel in
terms of conformal eigenfunctions. In this space of functions,
the BFKL kernel also enjoys the property of holomorphic
separability. Translating back to the momentum representation,
the kernel in the corresponding spectral form
acquires $\delta$ function - like pieces which, because of the
special properties of the impact factors, do not contribute to
physical amplitudes.

As mentioned above, the BFKL kernel, when applied to the color
octet wave function in momentum space, satisfies the bootstrap
condition, which reflects the s-channel consistency of the BFKL
calculation.  When transforming to transverse coordinates one
finds that, in the space of M\"obius functions, this bootstrap
condition cannot be satisfied, i.e. the colour octet wave
function lies outside this space of functions. In this paper we
will define a similarity transform, $\Phi$, which takes us from
the M\"obius space of functions (M space) to another space of
functions (named `analytic Feynman diagram' (AF) space) in which
the bootstrap holds. With the same transformation we can also
define `deformed' M\"obius transformations and a `deformed'
spectral representation of the BFKL kernel.

The paper is organized as follows. In the following section we review the
properties of the BFKL kernel, both in momentum space and in the space of
transverse coordinates. In section 3 we introduce the deformed
representation, and we define the similarity transform which takes us from
the M\"obius space ($M$ space) to the new deformed space ($AF$ space) of
functions. In section 4 we explicitly show that, in the $AF$ representation,
the bootstrap properties are fulfilled. Section 5 is devoted to the deformed
representation of the conformal algebra which follows from the
particular choice of the scalar product; in particular we compute
the transformed generators of the M\"obius group $\mathfrak{sl}(2,\mathbb{C})$.
Some details of our calculations are collected in two appendices.
%%%%%%%%%%%%%%%%%%%%%%%%%%%%%%%%%%%%%%%%%%%%%%%%%%%%%%%%%%%%%%%%%%%%%%%%%%

\section{The BFKL equation and bootstrap relation in LLA}

In this section we give a brief review of the BFKL approach describing the
dynamics of the reggeized gluons in LLA of perturbative QCD. Let us start
from the Schr\"odinger-like BFKL equation \cite{BFKL} describing the
compound state of two reggeized gluons,
\begin{equation}
K_2^{(R)} \otimes \psi_E = E \, \psi_E, \quad K_2^{(R)}= -\frac{N_c}{2}%
\left( \tilde{\omega}_1+ \tilde{\omega}_2 \right) - \lambda_R \bar{V}_{12}\,.
\label{bfkleq}
\end{equation}
Here $R$ labels the colour representation of the two gluon state and in the
singlet and octet channel one has respectively $\lambda_1=N_c$ and $%
\lambda_8=N_c/2$. The symbol $\otimes$ denotes an integration in the
transverse space in the case of the integral operator $\bar{V}_{12}$, while $%
E=-\omega=1-j$ where $j$ is the $t$-channel angular momentum. The
eigenvalues of $K_2^{R}$ give the positions of singularities of the $t$
channel partial waves, related to the scattering amplitude by the
Mellin transformation in the variable $\ln s$, where $s$ is the
squared total energy of colliding particles.

In LLA, in the momentum representation, the gluon trajectory (scaled by $%
N_c/2$) is given by the well known expression
\begin{equation}
\tilde{\omega}_i=\tilde{\omega}(\mbox{\boldmath $k$}_i)=- c \int d^2%
\mbox{\boldmath $l$} \, \frac{\mbox{\boldmath $k$}_i^2}{\mbox{\boldmath $l$}%
^2 (\mbox{\boldmath $k$}_i-\mbox{\boldmath $l$})^2}, \quad c=\frac{g^2}{%
(2\pi)^3}  \label{gluontrajectory}
\end{equation}
and the interaction term is defined by its action on the wave functions
\begin{eqnarray}
&& \bar{V}_{12} \otimes \phi \, (\mbox{\boldmath $k$}_1,\mbox{\boldmath $q$}-%
\mbox{\boldmath $k$}_1) = \int d^2 \mbox{\boldmath $k$}_1^{\prime}\, \bar{V}(%
\mbox{\boldmath $k$}_1, \mbox{\boldmath $q$}-\mbox{\boldmath $k$}_1| %
\mbox{\boldmath $k$}_1^{\prime},\mbox{\boldmath $q$} -\mbox{\boldmath $k$}%
_1^{\prime}) \phi(\mbox{\boldmath $k$}_1^{\prime},\mbox{\boldmath $q$} -%
\mbox{\boldmath $k$}_1^{\prime}) \\
&=& c \int d^2 \mbox{\boldmath $k$}_1^{\prime}\, \left[ \frac{%
\mbox{\boldmath $k$}_1^2}{
(\mbox{\boldmath $k$}_1^{\prime})^2 (\mbox{\boldmath
$k$}_1-\mbox{\boldmath $k$}_1^{\prime})^2} + \frac{(\mbox{\boldmath $q$}-%
\mbox{\boldmath $k$}_1)^2}{(\mbox{\boldmath $q$}-\mbox{\boldmath $k$}%
_1^{\prime})^2 (\mbox{\boldmath $k$}_1-\mbox{\boldmath $k$}_1^{\prime})^2} -
\frac{\mbox{\boldmath $q$}^2}{
(\mbox{\boldmath $k$}_1^{\prime})^2 (%
\mbox{\boldmath $q$}-\mbox{\boldmath $k$}_1^{\prime})^2} \right] \phi(%
\mbox{\boldmath $k$}_1^{\prime},\mbox{\boldmath $q$}-\mbox{\boldmath $k$}%
_1^{\prime})  \nonumber
\end{eqnarray}
and $\mbox{\boldmath $q$}=\mbox{\boldmath $k$}_1+\mbox{\boldmath $k$}_2$.
Let us note that this form is proper when the kernel acts on amputated
(without gluon propagators) impact factors, otherwise the propagators should
be removed from the kernel, as for example is the case for the expression in
eq. (\ref{BFKL_mom}).

We stress that this form of the LL BFKL kernel is obtained directly from the
Feynman diagrams analysis in perturbative QCD and has a simple analytic
behavior in the transverse momentum space. In particular it does not contain
$\delta$-functions $\delta ^2(k^{\prime}_1)$ and $\delta ^2(q-k^{\prime}_1)$.

In the construction of the BFKL kernels a very important property is the
gluon reggeization which can be verified with the bootstrap relation for the
amplitude with the octet quantum numbers \cite{BFKL}. The bootstrap relation
is a consequence of the $s$-channel unitarity. It claims, that the
scattering amplitude with octet quantum numbers obtained as a solution of
the corresponding BFKL equation should coincide with the Born term
multiplied by the Regge factor $s^{\tilde{\omega}(\bq)}$.

Because the BFKL equation was obtained by summing contributions from the
multi-particle production described by the multi-Regge amplitudes, the
bootstrap relation valid in the leading and next-to-leading orders allows to
connect the gluon trajectory and interaction terms. In LLA the bootstrap
condition for the BFKL kernel can be written as
\begin{equation}
\omega(\mbox{\boldmath $q$})-\omega(\mbox{\boldmath $k$}_1)-\omega(%
\mbox{\boldmath $k$}_2)=\bar{V}_{12} \otimes 1 \quad \mathrm{{or} \quad \bar{%
K}_{12}^{(8)}\otimes 1=-\omega(\mbox{\boldmath $q$})\, ,}
\end{equation}
where the constant $1$ is the wave function which can be conveniently
obtained after rescaling any function depending only on $\mbox{\boldmath $q$}
$, and the kernel $K_{12}^{(8)}$ acts on the amputated amplitudes. Let us
note that it is crucial in the bootstrap relation in LLA that the two gluons
are located at the same point in the transverse coordinate plane. It has
played a crucial role in the discovery of the leading Odderon solution in
the LLA of perturbative QCD~\cite{BLV} and also some general relation~\cite
{vacca} between bound states of $n$ and $n+1$ reggeized gluons.

In the following we shall use its equivalent form~\cite{vacca}, which has
the virtue of being infrared finite and which uses the standard BFKL singlet
kernel:
\begin{equation}
\bar{K}_{12}^{(1)}\otimes 1=-2 \omega(\mbox{\boldmath $q$}) +\omega(%
\mbox{\boldmath $k$}_1)+\omega(\mbox{\boldmath $k$}_2)\,.  \label{bootstrap2}
\end{equation}

On using an infrared mass regularization ($m \to 0$) one may write
\begin{equation}
\omega(\mbox{\boldmath $k$}) = - \frac{1}{2} \bar{\alpha}_s \log \left(
\frac{\mbox{\boldmath $k$}^2}{m^2} \right) \,,
\end{equation}
where $\bar{\alpha}_s=\alpha_s N_c/\pi$ and $\alpha_s=g^2/(4\pi)$, so that a
relation equivalent to the bootstrap, but explicitely infraredly
finite, is given by
\begin{equation} \bar{K}_{12}^{(1)}\otimes 1
= \frac{1}{2} \bar{\alpha}_s \log\left( \frac{ \mbox{\boldmath
$q$}^4}{\mbox{\boldmath $k$}_1^2 \mbox{\boldmath $k$}_2^2}
\right) \, .  \label{bootstrapIRfinite}
\end{equation}

The 2-gluon kernel (\ref{bfkleq}) in the singlet channel has been
investigated in great details in the coordinate representation~\cite
{Lipatov86}. The amplitude for the scattering of colorless objects is
factorized, in the high energy limit, in the product of the Green's function
(which exponentiates the BFKL kernel) and two  impact
factors. The impact factors vanish as the momentum of one of the
attached reggeized gluons goes to zero.
This permits to choose a special representation for the two-gluon propagator
and for the full BFKL kernel acting in the space of the so called M\"obius
functions (for two gluon states these are functions $f(\rho_1,\rho_2)$ such
that $f(\rho,\rho)=0$)~\cite{BLV2}. This is a special choice among infinite
others compatible with the gauge freedom of integrating a colorless impact
factor with any function belonging to the set defined by the equivalence
relation
\begin{equation}
f(\mbox{\boldmath $\rho $}_{1},\mbox{\boldmath $\rho $}_{2}) \sim \tilde{f}(%
\mbox{\boldmath $\rho $}_{1},\mbox{\boldmath $\rho $}_{2}) = f(%
\mbox{\boldmath $\rho $}_{1},\mbox{\boldmath $\rho $}_{2})+ f^{(1)}(%
\mbox{\boldmath $\rho $}_{1})+f^{(2)}(\mbox{\boldmath $\rho $}_{2})\,,
\label{uv_transf}
\end{equation}
since the corresponding shift in the momentum representation is proportional
to $\delta^{(2)}(\mbox{\boldmath $p$}_i)$.

In such a representation the operators are M\"obius (conformal) invariant~
\cite{Lipatov86}. Using complex coordinates, the BFKL hamiltonian
$H_{12}=K_{12}^{(1)}\,2 /\bar{\alpha}_s $ acting on
functions with propagators included can be written in the operator form \cite
{int}
\begin{equation}
H_{12}=\ln \,\left| p_{1}\right| ^{2}+\ln \,\left| p_{2}\right| ^{2}+\frac{1%
}{p_{1}p_{2}^{\ast }}\ln \,\left| \rho _{12}\right| ^{2}\,p_{1}p_{2}^{\ast
}\,+\frac{1}{p_{1}^{\ast }p_{2}}\ln \,\left| \rho _{12}\right|
^{2}\,p_{1}^{\ast }p_{2}-4\Psi (1)\,,  \label{BFKL_mom}
\end{equation}
where $\Psi (x)=d\ln \Gamma (x)/dx$, and we introduced the gluon holomorphic
momenta $p_r=i\partial /\partial \rho _r$. On the M\"obius space of
functions the holomorphic separability applies:
\begin{equation}
H_{12}=h_{12}+h_{12}^{\ast },\,\,h_{12}=\sum_{r=1}^{2}\left( \ln p_{r}+\frac{%
1}{p_{r}}\ln (\rho _{12})\,p_{r}-\Psi (1)\right) \,,  \label{BFKL_sep}
\end{equation}
and it is possible to find more easily the solutions of the homogeneous BFKL
pomeron equation which belong to irreducible unitary representations of the
M\"obius group. In particular the symmetry generators in this representation
are
\begin{equation}
M_{r}^{3}=\rho _{r}\partial _{r}\,,\,\,M_{r}^{+}=\partial
_{r}\,,\,\,M_{r}^{-}=-\rho^2_{r}\partial _{r}\, .  \label{moebius_gen}
\end{equation}
For two reggeized gluons one has $M^k=\sum_{r=1}^2 M^k_r$ and the Casimir
operator is defined as follows
\begin{equation}
M^2 = |\vec{M}|^2=-\rho _{12}^2\,\partial _1\,\partial _2\,,
\label{moebius_cas}
\end{equation}
where $\vec{M}=\sum_{r=1}^{2}\vec{M}_{r}$ and $\vec{M}_r%
\equiv(M_r^+,M_r^-,M_r^3)$.

The eigenfunctions of the BFKL kernel are also eigenstates of two Casimir
operators and given by \cite{Lipatov86}
\begin{equation}
E_{h,\bar{h}}(\brho_{10},\, \brho_{20}) \equiv \langle \rho | h \rangle =\left(%
\frac{ \rho _{12}}{\rho _{10}\rho _{20}}\right) ^{h}\left( \frac{\rho
_{12}^{*}}{ \rho _{10}^{*}\rho _{20}^{*}}\right) ^{\bar{h}}\,,
\label{pomstatescoord}
\end{equation}
where $h=\frac{1+n}{2}+i\nu$ ,\,\,$\bar{h}=\frac{1-n}{2}+i\nu$ are conformal
weights for the principal series of the unitary representations of the
M\"obius group, $n$ is the conformal spin and $d=1-2i\nu $ is the anomalous
dimension of the operator $O_{h,\bar{h}}(\mbox{\boldmath $\rho $}_{0})$
describing the compound state of two reggeized gluons (note, that here and
below we use other notations for conformal weights $m,\widetilde{m}$ in
Refs. \cite{Lipatov86, Lcft}). The corresponding eigenvalues of the BFKL
kernel are given by
\begin{equation}
\chi_h \equiv \chi(\nu,n)=\bar{\alpha}_s \left( \psi(\frac{1+|n|}{2}+i\nu)
+\psi(\frac{1+|n|}{2}-i\nu) -2\psi(1)\right) = \bar{\alpha}_s \,
\epsilon_h\, .  \label{eigenvalues}
\end{equation}

The action of the BFKL kernel can also be written, again on the space of
M\"obius functions, after a duality transformation~\cite{int, dual, BLV2},
in terms of integral operator appearing in the dipole picture \cite{dipole}
\begin{equation}
H_{12}\,f_{\omega }(\mbox{\boldmath $\rho$}_{1},\mbox{\boldmath $\rho $}%
_{2})= \int \frac{d^{2}\brho _{3}}{\pi }\,\frac{\left| \brho _{12}\right| ^{2}%
}{\left| \brho _{13}\right| ^{2}\left| \brho _{23}\right| ^{2}}\,\left(
f_{\omega }(\mbox{\boldmath $\rho $}_{1},\mbox{\boldmath $\rho $}%
_{2})-f_{\omega }(\mbox{\boldmath $\rho $}_{1},\mbox{\boldmath $\rho $}%
_{3})-f_{\omega }(\mbox{\boldmath $\rho $}_{2},\mbox{\boldmath $\rho $}%
_{3})\right) \,.
\end{equation}

We now consider the pomeron eigenstates in the momentum representation.
Different forms of the Fourier transform of the function in eq. (\ref
{pomstatescoord}) have been given in the most general case. One finds a sum
of an analytic term $\tilde{E}_{h,\bar{h}}^A$, which has been given in an
explicit way or in a integral form, and a $\delta$-like one $\tilde{E}_{h,%
\bar{h}}^{\delta}$. More precisely we obtain
\begin{eqnarray}
\tilde{E}_{h,\bar{h}}(\mbox{\boldmath $k$}_1,\mbox{\boldmath $k$}_2)&=& \int{%
\frac{d^2\mbox{\boldmath $r$}_1}{(2\pi)^2}}{\frac{d^2\mbox{\boldmath $r$}_2
}{(2\pi)^2}}\; \left({\frac{r_{12}}{r_1r_2}}\right)^h \left({\frac{r^*_{12}}{%
r^*_1r^*_2}} \right)^{\bar{h}} e^{i(\mbox{\boldmath $k$}_1\cdot%
\mbox{\boldmath $r$}_1+\mbox{\boldmath $k$}_2\cdot\mbox{\boldmath $r$}_2)}=
\tilde{E}_{h,\bar{h}}^{A}(\mbox{\boldmath $k$}_1,\mbox{\boldmath $k$}_2)+
\tilde{E}_{h,\bar{h}}^{\delta}(\mbox{\boldmath $k$}_1,\mbox{\boldmath $k$}_2)
\nonumber \\
&=& \langle k | h \rangle=\langle k | h^A \rangle+\langle k | h^\delta
\rangle \,,  \label{pom_mom}
\end{eqnarray}
where a bra-ket compact notation is also introduced. By explicit computation
one finds
\begin{equation}
\tilde{E}^{\delta}_{h,\bar{h}}(\mbox{\boldmath $k$}_1,\mbox{\boldmath $k$}%
_2)= \Bigl[ \delta^{(2)}(\mbox{\boldmath $k$}_1) +(-1)^n \delta^{(2)}(%
\mbox{\boldmath $k$}_2) \Bigr] \frac{i^n}{2\pi} 2^{1-h-\bar{h}} \frac{%
\Gamma(1-\bar{h})}{\Gamma(h)} q^{\bar{h}-1} q^{*\, h-1 } \,,
\label{pom_delta}
\end{equation}
where the complex form $v=v_x+i v_y$ of any bidimensional transverse vector $%
\mbox{\boldmath $v$}=(v_x,v_y)$ (coordinate or momentum) is used. In the
calculation of a physical cross section one performs integrations with
two-gluon colourless impact factors and this $\delta$-like term gives zero
contribution, so that all the physics is related to the analytic term. In
fact the $\delta$-terms have been introduced when one has chosen to move to
the M\"obius representation which is characterized by the simple and
beautiful form in the coordinate representation.

The analytic part can be written, for example, as given in~\cite{BBCV}
wherein one finds
\begin{equation}
\tilde{E}_{h\bar{h}}^{A}(\mbox{\boldmath $k$}_1,\mbox{\boldmath $k$}_2)=C_v%
\Big(X(\mbox{\boldmath $k$}_1,\mbox{\boldmath $k$}_2)+ (-1)^nX(%
\mbox{\boldmath $k$}_2,\mbox{\boldmath $k$}_1)\Big),  \label{pom_an}
\end{equation}
with the coefficient $C_v$ given by
\begin{equation}
C_v=\frac{(-i)^n}{(4\pi)^2}h\bar{h} (1-h)(1-\bar{h})\Gamma(1-h)\Gamma(1-\bar{%
h}).  \label{fullC}
\end{equation}
The expression for $X$ in complex notation was written in terms of the
hypergeometric functions
\begin{eqnarray}
X(\mbox{\boldmath $k$}_1,\mbox{\boldmath $k$}_2)=\left(\frac{k_1}{2}\right)^{%
\bar{h}-2} \left(\frac{k^*_2}{2}\right)^{h-2} {}_2F_1
\left({
\begin{array}{cc}
1-h,2-h\\
2
\end{array}
}\bigg|-\frac{k^*_1}{k^*_2}\right)
{}_2F_1 \left({
\begin{array}{cc}
1-\bar{h},2-\bar{h}\\
2
\end{array}
}\bigg|-\frac{k_2}{k_1}\right) \,.   \label{fullX}
\end{eqnarray}
Another expression for $\tilde{E}_{h\bar{h}}^{A}(\mbox{\boldmath $k$}_1,%
\mbox{\boldmath $k$}_2)$ which will be very useful for our purposes has been
given in~\cite{LV1} in an integral form:
\begin{equation}
\tilde{E}_{h\bar{h}}^{A}(\mbox{\boldmath $k$}_1,\mbox{\boldmath $k$}_2)= C_l
\frac{1}{\mbox{\boldmath $k$}_1^2 \mbox{\boldmath $k$}_2^2} \int d^2 %
\mbox{\boldmath $p$} \left[ \frac{p\, (q-p)}{k_1-p}\right]^{\bar{h}-1} \left[
\frac{p^* (q^*-p^*)}{k_1^*-p^*}\right]^{h-1} \, ,  \label{ana_lip}
\end{equation}
where
\begin{equation}
C_l=-(-1)^n \frac{i^{\bar{h}-h} h \, \bar{h}}{2^{h+\bar{h}} \pi^3} \frac{%
\Gamma(1-h)}{\Gamma(\bar{h})}\, .  \label{Cl}
\end{equation}

Below we shall consider the completeness relation and the spectral
representation for the BFKL kernel (cf. Ref.~\cite{Lcft}).

Let us before discuss the scalar product in the space of functions where the
kernel is acting on. For the kernel in the momentum space associated to the
Feynman diagram derivation there is only one scalar product available. If we
consider non amputated functions of two momenta $\mbox{\boldmath $k$}_1,%
\mbox{\boldmath $k$}_2$, i.e. with 2-dimensional propagators $1/(%
\mbox{\boldmath $k$}_1^2 \mbox{\boldmath $k$}_2^2)$ included, the scalar
product is defined as
\begin{equation}
\langle f | g \rangle \equiv \int d\mu(\mbox{\boldmath $k$}) f^*(%
\mbox{\boldmath $k$}_1,\mbox{\boldmath $k$}_2) g(\mbox{\boldmath $k$}_1,%
\mbox{\boldmath $k$}_2)\,,
\end{equation}
with the integration measure
\begin{equation}
d \mu(\mbox{\boldmath $k$}) =\mbox{\boldmath $k$}_1^2 \mbox{\boldmath $k$}%
_2^2 \, \,\delta^2(\mbox{\boldmath $q$}-\mbox{\boldmath $k$}_1-%
\mbox{\boldmath $k$}_2) \, d^2\mbox{\boldmath $k$}_1 d^2\mbox{\boldmath $k$}%
_2  \label{scal_prod}
\end{equation}
which kills the extra propagators.

In the M\"obius space of functions (in coordinate space) for the principal
series of the M\"{o}bius group representation, whenever functions with
conformal weight $h=0$ or $h=1$ are not considered, two possible choices are
available ~\cite{int}. Since the M\"obius functions contain
propagators, one possibility is to remove, as before, the
propagators in one function, acting with the operator
$\mbox{\boldmath $\partial$}_1^2 \mbox{\boldmath
$\partial$}_2^2$.  The other possibility, related to the form of
the Casimir operator of the M\"obius group, is to use instead the
measure $d^2\mbox{\boldmath $\rho$}_1 d^2\mbox{\boldmath
$\rho$}_2 /|\rho_{12}|^4$. The latter is not equivalent to the
former when we consider the conformal wheights $h=0$ or $h=1$.
Note, that, providing that we go from the Regge kinematics to the
deep-inelastic scattering in the Bjorken regime, the additional
series of the unitary representations of the M\"{o}bius group
should be also used ~\cite{LV1} (cf. Ref. ~\cite{DKKM}).

We now consider the completeness relation in coordinate space~\cite
{Lipatov86}, in the space of M\"obius functions, where the eigenfunctions of
the Casimir operator of the M\"obius group, defined in (\ref{pomstatescoord}%
), constitute a suitable spectral basis for the operators acting on the
space of the colourless impact factors, which, we stress, never ``feel'' the
presence of the terms with a $\delta$-distribution behavior in momentum
space, i.e. those of eq. (\ref{pom_delta}). For the operators amputated
(spatial propagators removed) to the left (with notation $\hat{1}_L$ and $%
\hat{\bar{K}}_{12}^{(1)}$) one can write  \cite{Lipatov86}
\begin{eqnarray}
\!\!\!\!\! \langle \rho |\hat{1}_L| \rho^{\prime}\rangle \!\!\! &\equiv&
\!\!\!(2\pi)^4 \delta^2(\brho_{11^{\prime}})\, \delta^2(\brho_{22^{\prime}}) =
\! \int d^2 \!\mbox{\boldmath $\rho$}_0 \sum_h \frac{N_h}{|\brho_{12}|^4} \,
E_{h,\bar{h}}(\brho_{10},\, \brho_{20})\, E_{h,\bar{h}}^{*}(\brho_{1^{%
\prime}0},\, \brho_{2^{\prime}0}) \,,  \nonumber \\
\!\!\!\!\langle \rho |\hat{\bar{K}}_{12}^{(1)}| \rho^{\prime}\rangle
\!\!\!\!&=& \!\!\! \int d^2 \!\mbox{\boldmath $\rho$}_0 \sum_h \frac{N_h}{%
|\brho_{12}|^4} \, E_{h,\bar{h}}(\brho_{10},\, \brho_{20})\, \chi_h \,E_{h,\bar{%
h}}^{*}(\brho_{1^{\prime}0},\, \brho_{2^{\prime}0}) \,,
\label{coord_relations}
\end{eqnarray}
where we have defined the weights $N_h$, as well as $\tilde{N_h}$ for later
use,
\begin{equation}
N_h= 16(\nu^2+n^2/4) \, , \quad \tilde{N}_h=(2\pi)^2 \frac{\nu^2+n^2/4}{%
[\nu^2+(n-1)^2/4][\nu^2+(n+1)^2/4]}
\end{equation}
and $\sum_h \equiv \sum_n \int d\nu$. If we try to extrapolate these
operators to a wider domain which contains functions which no longer vanish
for zero momenta, contrary to the colorless impact factor case, one needs
some care. For example, for the  function, which in
momentum space depends only on the total momentum, just the case
we meet in the bootstrap relation in LLA, one obtains the
the result $\delta^2(\rho_{12})$  in coordinate space
(for the amputated impact factor). But such generalized function
is orthogonal to the basis constituted by the functions in eq.
(\ref {pomstatescoord}) since
$\int d^2\!\mbox{\boldmath
$\rho$}_{1^{\prime}} E_{h, \bar{h}}^{*}(\rho_{1^{\prime}0},\,
\rho_{2^{\prime}0}) \delta^2(\rho_{1^{\prime}2^{\prime}}) =0$ and
therefore the application of the operators in eq.
(\ref{coord_relations}) on this function will give zero. In other
words the bootstrap relation cannot be fullfilled in such a
framework with the spectral representations given in eq. (\ref
{coord_relations}).

But this should not be surprising, since the choice done thanks to the
equivalence relation in eq. (\ref{uv_transf}), is based on a restriction of
the action of the kernel to the functions similar to colorless
impact factors.
%%%%%%%%%%%%%%%%%%%%%%%%%%%%%%%%%%%%%%%%%%%%%%%%%%%%%%%%%%%%%%%%%%%%

\section{Deformed representation}

Moving to the momentum representation, one may write the Fourier transform
of the relations in (\ref{coord_relations}) and the same considerations done
above may be repeated in this case, leading to the same conclusions. In
particular one cannot write a bootstrap relation in a spectral basis built
upon the states of eq. (\ref{pom_mom}). On the other hand one notices that
the BFKL kernel $K_{12}^{(1)}$ in momentum space, as constructed from the
Feynman diagram analysis, has an analytic behavior which must be reflected
also in its spectral decomposition. It can also be noted that the behavior
in $|\mbox{\boldmath $q$}|$ of the $\delta$-like term (\ref{pom_delta}) is $|%
\mbox{\boldmath $q$}|^{-1+2i\nu}$, which is singular in the forward limit
where analytic behavior is expected for fixed $\vec{k}_1=-\vec{k}_2$.

Therefore it is natural to suggest the modified relations for the
eigenfunction completeness and for the spectral representation of the BFKL
kernel:
\begin{eqnarray}
\langle k |\hat{1}_L| k^{\prime}\rangle &\equiv& 1 = \sum_h \tilde{N}_h \,
\tilde{E}_{h\bar{h}}^{A}(\mbox{\boldmath $k$}_1,\mbox{\boldmath $k$}_2) \,
\tilde{E}_{h\bar{h}}^{A*}(\mbox{\boldmath $k$}_1^{\prime},\mbox{\boldmath
$k$}_2^{\prime}) \,,  \nonumber \\
\langle k |\hat{\bar{K}}_{12}^{(1)}| k^{\prime}\rangle &=& \sum_h \tilde{N}%
_h \, \tilde{E}_{h\bar{h}}^{A}(\mbox{\boldmath $k$}_1,\mbox{\boldmath $k$}%
_2) \, \chi_h \,\tilde{E}_{h\bar{h}}^{A*}(\mbox{\boldmath $k$}_1^{\prime},%
\mbox{\boldmath $k$}_2^{\prime})\, \,,  \label{mom_relations} \\
\langle k |\hat{G}_{12}^{(1)}(y)| k^{\prime}\rangle &=& \sum_h \tilde{N}_h
\, \tilde{E}_{h\bar{h}}^{A}(\mbox{\boldmath $k$}_1,\mbox{\boldmath $k$}_2)
\, e^{y \chi_h} \,\tilde{E}_{h\bar{h}}^{A*}(\mbox{\boldmath $k$}_1^{\prime},%
\mbox{\boldmath $k$}_2^{\prime})\,,  \nonumber
\end{eqnarray}
where only the analytic contribution from each state of the spectral basis
is used and the measure for integration is defined in eq. (\ref{scal_prod}),
according to our choice of considering functions with propagators included.
Below we shall verify this anzatz.

Firstly we consider a space of functions (with removed propagators) which
includes functions corresponding to the colorless impact factors and at
least one function depending only on the total momentum, which is
associated to a particular colored impact factor. Secondly we
modify the spectral representation of the LL BFKL kernel, wherein
the eigenvalues are assumed to be the same expressions of
the conformal weights as for the restriction on the M\"obius
space of function, and the eigenfunctions are deformed in order
to be analytic in the momentum space. A similar idea was already
considered in the literature in an attempt to couple the LL BFKL
pomeron to a quark~\cite{MT,BFLLRW}.

We proceed now with the definition of a transformation between the M\"obius
(M) basis and the analytic Feynman (AF) basis. Let us observe that one can
go from one basis to the other by a simple transformation, which reads in
coordinate representation as follows

\begin{eqnarray}
\Phi^{-1} :\mathrm{M \to AF}\!\!\!\! &,&\!\!\!
E^A_h(\brho_{10},\brho_{20})=E^M_h(\brho_{10},\brho_{20})-
\lim_{\rho_1\to\infty} E^M_h(\brho_{10},\brho_{20})-\lim_{\rho_2\to\infty}
E^M_h(\brho_{10},\brho_{20})  \nonumber \\
&&=E^M_h(\brho_{10},\brho_{20})-2^{-h-\bar{h}}
\left(E^M_h(-\brho_{20},\brho_{20})+(-1)^n
E^M_h(-\brho_{10},\brho_{10})\right)\,,  \label{mapping_on_E} \\
\Phi :\mathrm{AF \to M :}\!\!\!\! &,&\!\!\!
E^M_h(\brho_{10},\brho_{20})=E^A_h(\brho_{10},\brho_{20})+\frac{1}{2^{h+\bar{h}%
}-2} \left( E^A_h(-\brho_{20},\brho_{20})+(-1)^n
E^A_h(-\brho_{10},\brho_{10})\right) \,.  \nonumber
\end{eqnarray}
These relations are clearly non local and have been obtained by noting that
\begin{equation}
E^M_h(-\brho,\brho)=2^{h+\bar{h}} {\left( \frac{1}{\rho} \right)}^h {\left(%
\frac{1}{\rho^{*}}\right)}^{\bar{h}} \, , \quad E^A_h(-\brho,\brho)=(2^{h+\bar{%
h}} -2){\left( \frac{1}{\rho} \right)}^h {\left(\frac{1}{\rho^{*}}\right)}^{%
\bar{h}} \, .
\end{equation}
For the case of even conformal spins ($n$) the second transformation can be
written in a simple local form on observing that
\begin{equation}
\frac{1}{2^{h+\bar{h}}-2} \left( E^A_h(-\brho_{20},\brho_{20})+(-1)^n
E^A_h(-\brho_{10},\brho_{10})\right)= - \frac{1}{2}\left(E^A_h(\brho_{10},%
\brho_{10})+(-1)^n E^A_h(\brho_{20},\brho_{20}) \right)
\end{equation}
while for odd conformal spin such a simple transformation is not possible
since $E^A_h(\rho,\rho)=0$. It is related to the fact, that in the last case
both functions $E^M_h(\rho_{10},\rho_{20})$ and $E^A_h(\rho_{10},\rho_{20})$
satisfy the colour transparancy property $E^{M,A}_h(\rho,\rho)=0$, but only $%
E^M_h(\rho_{10},\rho_{20})$ has the simple conformal properties.

It can be easily checked that
\begin{equation}
\Phi \Phi^{-1} \equiv I_M \, , \quad \Phi^{-1} \Phi \equiv I_{AF}
\label{inverse}
\end{equation}
so that $\Phi$ is a 1-1 mapping with inverse really given by $\Phi^{-1}$.

The Green's function for the evolution in the rapidity $y$ based on the
deformed analytic spectral basis was also written above (see (\ref
{mom_relations})). In the next section we shall show that with this
prescription one is recovering the correct relation compatible with the
bootstrap requirement for the gluon reggeization in LLA.

\begin{figure}[ht]
\begin{center}
\includegraphics[width=7cm]{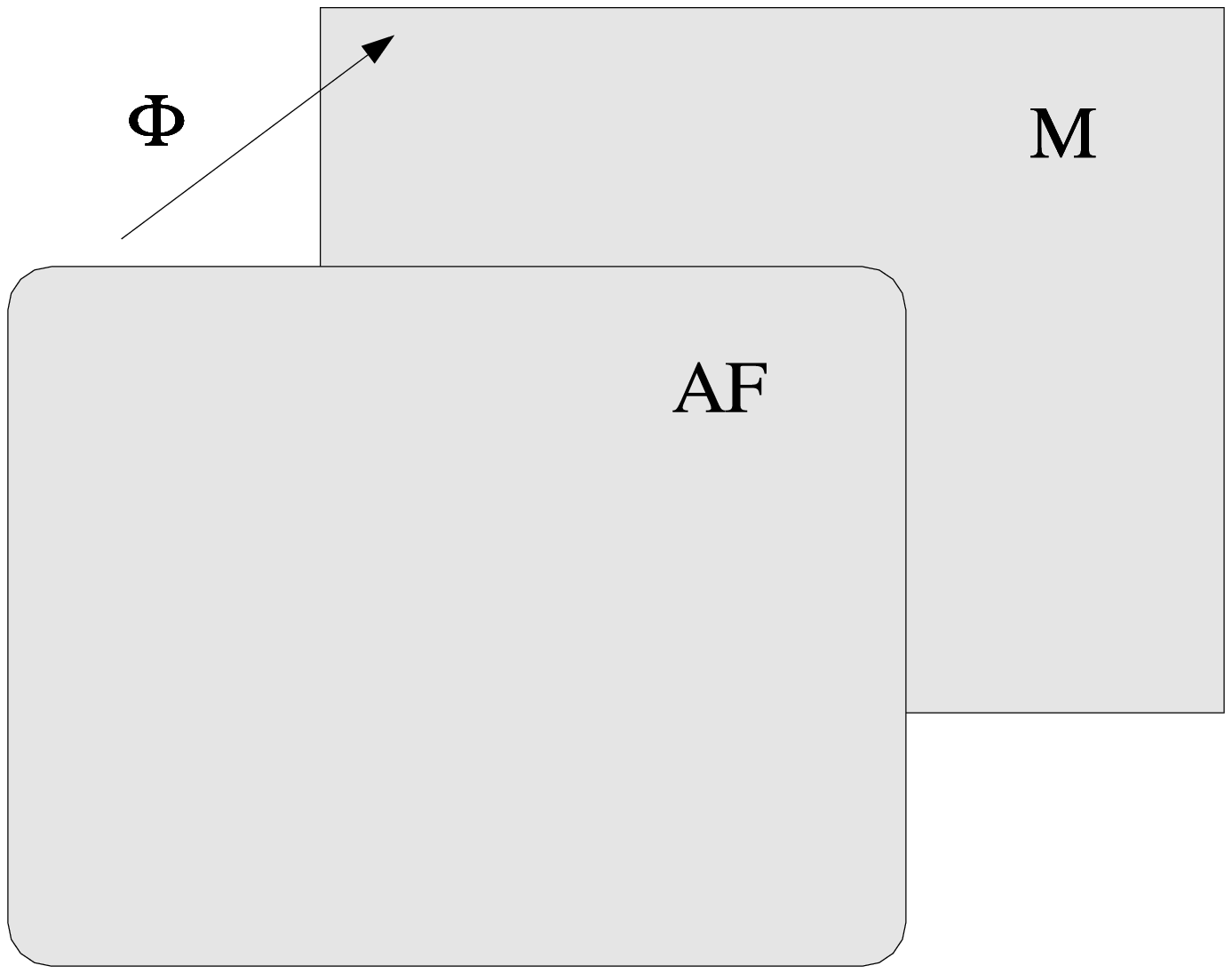}
\end{center}
\caption{There is a $1$-$1$ mapping between the AF-space and the M-space.}
\label{fig_mapping}
\end{figure}

Since the spectrum in both representations is the same it is natural to
expect that the BFKL kernel satisfying the bootstrap relation is conformal
invariant, in the new deformed analytic Feynman basis obtained by a
similarity transformation given by the operator $\Phi$. Namely, the well
known generators of the M\"{o}bius group $\vec{M}_r$ could be extended to
this basis as follows
\begin{equation}
\vec{M}^{AF}_r=\Phi^{-1} \vec{M}_r \Phi \,.  \label{deREP}
\end{equation}
Let us summarize the result. If one considers the impact factors derived
from Feynman diagrams calculations, they are the functions belonging to the
AF-space and not to the M-space. We remind that we have two different
completeness relations based on the two spectral basis $E^{AF}_h$ and $E_h^M$
and in these two spaces there are different definitions for scalar products
and normalizations of wave functions. Nevertheless the cardinality of the
two basis is the same. The two spaces are related by the 1-1 mapping {$\Phi:$
AF-space $\to$ M-space} as previously discussed (see also Fig. \ref
{fig_mapping}) and one space is \textit{shifted} with respect to the other.
Moreover any operator $O^{AF}$ in the AF-space which can be decomposed on
the spectral basis can be defined on the M-space as $O^M=\Phi O^{AF}
\Phi^{-1}$ and viceversa $O^{AF}=\Phi^{-1} O^{M} \Phi$. This can be seen as
an isometry between the M\"obius space and the analytical Feynman space with
the different scalar products in these spaces.

%figure--------- The mapping  ----------------------
%figure--------------------------------------------------

The next interesting question is therefore if it is possible to construct
explicitely the new representation defined in eq. (\ref{deREP}). We shall
study this problem in the section $5$, starting from analyzing the Fourier
transform of the M\"obius algebra in the M-space and later on moving to the
AF-space. In the next section we will show that with the meromorphic
prescription we indeed recover the bootstrap properties of the BFKL kernel.
%%%%%%%%%%%%%%%%%%%%%%%%%%%%%%%%%%%%%%%%%%%%%%%%%%%%%%%%%%%%%%%%%%%%%%%%%%

\section{AF representation and the bootstrap relation}

For the sake of using a compact notation, let us introduce for some
functions of transverse gluon momenta the corresponding wave functions for a
color singlet state. These are the \textit{power} $|P_\lambda\rangle$, the
\textit{unity} $|U\rangle$ and the \textit{log} $|L\rangle$ states (modulus
the propagators), defined by their representations in the momentum space:
\begin{eqnarray}
\langle k | P_\lambda \rangle &\equiv& \frac{1}{\mbox{\boldmath $k$}_1^2 %
\mbox{\boldmath $k$}_2^2} \left( \frac{\mbox{\boldmath $q$}^4}{%
\mbox{\boldmath $k$}_1^2 \mbox{\boldmath $k$}_2^2} \right)^ \lambda \, ,
\nonumber \\
\langle k | U \rangle &\equiv& \frac{1}{\mbox{\boldmath $k$}_1^2 %
\mbox{\boldmath $k$}_2^2}= \left( \langle k | P_\lambda \rangle
\right)_{\lambda=0} \, ,  \nonumber \\
\langle k | L \rangle &\equiv& \frac{1}{\mbox{\boldmath $k$}_1^2 %
\mbox{\boldmath $k$}_2^2} \log\left( \frac{\mbox{\boldmath $q$}^4}{%
\mbox{\boldmath $k$}_1^2 \mbox{\boldmath $k$}_2^2} \right) = \frac{d}{d
\lambda} \left( \langle k | P_\lambda \rangle \right)_{\lambda=0}\, .
\label{power_rel}
\end{eqnarray}
Since, as previoulsy discussed, one has for the M\"{o}bius-invariant wave
function $h$ the equality $\langle h | U \rangle=0$ we obtain
\begin{equation}
\langle h^A | U \rangle =- \langle h^\delta | U \rangle \, .
\label{ana_delta}
\end{equation}
Let us recall the completeness relation and the BFKL kernel written as
\begin{eqnarray}
\hat{1}_L &=& \sum_h \tilde{N}_h \, |h^{A}\rangle \,\langle h^{A} | \,,
\nonumber \\
\hat{\bar{K}}_{12}^{(1)}&=& \sum_h \tilde{N}_h \, |h^{A}\rangle \, \chi_h \,
\langle h^{A} | \, .  \label{gen_relations}
\end{eqnarray}
On using the previous relations we write the bootstrap relation
in the form of eq. (\ref{bootstrapIRfinite}) as
\begin{equation}
\bar{\alpha}_s \sum_h \tilde{N}_h \, |h^{A}\rangle \, \epsilon_h \, \langle
h^{A} | U \rangle = \frac{\bar{\alpha}_s}{2} \sum_h \tilde{N}_h \,
|h^{A}\rangle \,\langle h^{A} | L \rangle \, ,
\end{equation}
where eq. (\ref{eigenvalues}) was used. In order to show that our
definitions (\ref{gen_relations}) are correct, we should prove that the
coefficients of the conformal basis $|h^{A}\rangle$ on both the l.h.s. and
the r.h.s. do coincide, i.e.
\begin{equation}
\epsilon_h \, \langle h^{A} | U \rangle = \frac{1}{2}\langle h^{A} | L
\rangle \, .  \label{coeff_bootstrap}
\end{equation}
It is therefore enough to calculate the integral given by $\langle h^{A} |
P_\lambda \rangle$. Verifying that this expression is finite it is
sufficient also to calculate the first two terms of the Taylor expansion in $%
\lambda$ to verify the eq. (\ref{coeff_bootstrap}), thanks to eq. (\ref
{power_rel}).

We also note that the left hand side of eq.
(\ref{coeff_bootstrap}) can be calculated easily in an
independent way using the relations in eq. (\ref {ana_delta}) and
(\ref{pom_delta}), which give:
\begin{equation} \langle h^{A} |
U \rangle = -\delta_{n,\mathrm{even}} \frac{i^n}{2\pi} 2^{h+%
\bar{h}} \frac{\Gamma(h)}{\Gamma(1-\bar{h})}q^{* \,
-h}q^{-\bar{h}} \, , \label{hAU} \end{equation} where
$\delta_{n,\mathrm{even}}=[1+(-1)^n]/2$ has support (and value
$1$) for any even integer number.

Let us now find explicitely $\langle h^{A} | P_\lambda \rangle$.
It is convenient to perform the calculations in the momentum
space where we have defined the analytic element $| h^{A}
\rangle$ of the conformal basis.   Taking for it the
expression given in eq. (\ref{ana_lip}) we write:
\begin{eqnarray}
\langle h^{A} | P_\lambda \rangle &=& \int d^2
\mbox{\boldmath $k$}_1 d^2  \mbox{\boldmath $k$}_2 \,
\delta^{(2)}(\mbox{\boldmath $q$}-\mbox{\boldmath
$k$}_1-\mbox{\boldmath $k$}_2) \tilde{E}_{h\bar{h}}^{A\,
*}(\mbox{\boldmath $k$}_1,\mbox{\boldmath $k$}_2) \left(
\frac{\mbox{\boldmath $q$}^4}{ \mbox{\boldmath $k$}_1^2
\mbox{\boldmath $k$}_2^2} \right)^ \lambda \nonumber \\ &=& C_l^*
|\mbox{\boldmath $q$}|^{4\lambda} \int \! d^2 \mbox{\boldmath
$k$} \! \int \! d^2 \mbox{\boldmath $p$} \, p^{-\bar{h}}
(q-p)^{-\bar{h}} (k-p)^{ \bar{h}} \, k^{-1-\lambda}
(q-k)^{-1-\lambda} \times (\mathrm{h.c.}) \, .
\end{eqnarray}
Using new integration variables induced by the change $x=p/q$ and $y=q/k$
with analogous relations for the complex conjugated ones, we
obtains
\begin{equation}
\langle h^{A} | P_\lambda \rangle =
C_l^* q^{-\bar{h}} q^{*\, -h} I_\lambda(1) \, ,  \label{hapl}
\end{equation}
where
\begin{equation}
I_\lambda(z)= \int \! d^2 \mbox{\boldmath $x$} \, d^2 \mbox{\boldmath $y$}
\, x^{-\bar{h}} (1-x)^{-\bar{h}} y^{2 \lambda-\bar{h}} (1-y)^{-1-\lambda}
(1-x y z)^{\bar{h}} \times (h.c.) \, .  \label{integral}
\end{equation}
It is convenient to define the function $I_{\lambda}(z)$ because one can
find an explicit expression for it in terms of the generalized
hypergeometric functions ${}_3 F_2$. We illustrate some details of the
calculation in appendix A.

It is sufficient to represent the result of the integration in power series
of $\lambda$ for the point $z=1$:
\begin{equation}
I_\lambda(1)=\sum_m \frac{1}{m!} I^{(m)}_{\lambda}(1)\arrowvert_{\lambda=0}
\,\, \lambda^m \,.
\end{equation}

Recalling the first two terms in the expansion (see eq. (\ref{d01}) in
appendix A) one can see, on using the definitions of eq. ({\ref{power_rel})
and (\ref{hapl}), which imply $\langle h^{A} | U \rangle=(\langle h^{A} |
P_\la \rangle)_{\lambda=0}$ and $\langle h^{A} | L \rangle=(d\langle h^{A} |
P_\la \rangle/d\lambda)_{\lambda=0}$, that the relation in eq. (\ref
{coeff_bootstrap}) is verified and therefore also the bootstrap relation. As
a check one may evaluate explicitely the quantity
\begin{equation}
(\langle h^{A} | P_\la \rangle)_{\lambda=0}=C_l^* q^{-\bar{h}} q^{*\, -h}
\frac{2 \pi^2}{(1-h)(1-\bar{h})}\, \delta_{n,\mathrm{even}} \, ,
\label{even_check}
\end{equation}
using the eq. (\ref{Cl}). It is easy to show that it coincides with the
expression of eq. (\ref{hAU}), computed in a completely different way.%
\newline
In order to give an alternative check of the representation proposed, we
have also calculated explicitly the third term of the expansion of the BFKL
Green's function $G_{12}^{(1)}(y) = e^{y \bar{K}_{12}^{(1)}} \otimes 1$ (the
first term is trivial and the second corresponds to the bootstrap relation),
which contains the double iteration of the kernel acting on the
unity (momentum space). The details are in the appendix B. }

Using the spectral representation one can compute an arbitrary number of
iterations of the BFKL kernel or the action of the BFKL Green's function on
the impact factor of a quark or gluon line.
%%%%%%%%%%%%%%%%%%%%%%%%%%%%%%%%%%%%%%%%%%%%%%%%%%%%%%%%%%%%%%%%%%%%%%%%%%

\section{The deformed representation of the M\"obius (conformal) algebra}

In this section we will examine the representation of the
$\mathfrak{sl}(2,\mathbb{C})$ algebra
in the $AF$ space of funtions. To this end we start from the familiar
M\"obius algebra in the $M$ space and transform it
into momentum space. We then
switch to the $AF$ space: in order to retain the correct structure of the
algebra we have to construct a new representation of the
generator $M_{-}$.

Let us reconsider the generators of the conformal M\"obius group in $2$
dimensions in coordinate space given in eq. (\ref{moebius_gen}) and the
corresponding Casimir operator
 (\ref{moebius_cas}). We will be interested in
writing similar relations in momentum space. We start from the conformal
eigenfunctions (\ref{pomstatescoord}) which from now on will carry the
superscript '$M$', and we define their Fourier transforms:

\begin{eqnarray}
E_{h{\bar{h}}}^{M}(\mbox{\boldmath $\rho$}_{0};\mbox{\boldmath $\rho$}_{1},%
\mbox{\boldmath $\rho$}_{2}) &=&\Bigg(\frac{\rho _{12}}{\rho _{10}\rho _{20}}%
\Bigg)^{h}\Bigg(\frac{\rho _{12}^{\ast }}{\rho _{10}^{\ast }\rho _{20}^{\ast
}}\Bigg)^{{\bar{h}}}\,, \\
E_{h{\bar{h}}}^{M}(\mbox{\boldmath $q$};\mbox{\boldmath $k$}_{1},%
\mbox{\boldmath $k$}_{2})=\mathcal{F}[E_{h{\bar{h}}}^{M}](%
\mbox{\boldmath
$q$};\mbox{\boldmath $k$}_{1},\mbox{\boldmath $k$}_{2}) &=&\int d\mu ~e^{-i%
\mbox{\boldmath $q$}\cdot \mbox{\boldmath $\rho$}_{0}+i\mbox{\boldmath $k$}%
_{1}\cdot \mbox{\boldmath $\rho$}_{1}+i\mbox{\boldmath $k$}_{2}\cdot %
\mbox{\boldmath $\rho$}_{2}}E_{h{\bar{h}}}^{M}(\mbox{\boldmath $\rho$}_{0};%
\mbox{\boldmath $\rho$}_{1},\mbox{\boldmath $\rho$}_{2})=  \nonumber \\
&=&(2\pi )^{2}\delta ^{(2)}(\mbox{\boldmath $q$}-\mbox{\boldmath $k$}_{1}-%
\mbox{\boldmath $k$}_{2}){\tilde{E}}_{h{\bar{h}}}^{M}(\mbox{\boldmath $k$}%
_{1},\mbox{\boldmath $k$}_{2})\,, \\
{\tilde{E}}_{h{\bar{h}}}^{M}(\mbox{\boldmath $\rho$}_{0};\mbox{\boldmath $k$}%
_{1},\mbox{\boldmath $k$}_{2})=\tilde{\mathcal{F}}[E_{h{\bar{h}}}^{M}](%
\mbox{\boldmath $\rho$}_{0};\mbox{\boldmath
$k$}_{1},\mbox{\boldmath $k$}_{2}) &=&\int d\tilde{\mu}~e^{i%
\mbox{\boldmath
$k$}_{1}\cdot \mbox{\boldmath
$\rho$}_{1}+i\mbox{\boldmath $k$}_{2}\cdot \mbox{\boldmath
$\rho$}_{2}}E_{h{\bar{h}}}^{M}(\mbox{\boldmath
$\rho$}_{0};\mbox{\boldmath $\rho$}_{1},\mbox{\boldmath $\rho$}_{2})=
\nonumber \\
&=&e^{i(\mbox{\boldmath $k$}_{1}+\mbox{\boldmath $k$}_{2})\cdot
\mbox{\boldmath
$\rho$}_{0}}{\tilde{E}}_{h{\bar{h}}}^{M}(\mbox{\boldmath $k$}_{1},%
\mbox{\boldmath $k$}_{2})\,, \\
{\tilde{E}}_{h{\bar{h}}}^{M}(\mbox{\boldmath $k$}_{1},\mbox{\boldmath $k$}%
_{2})=\tilde{\mathcal{F}}[E_{h{\bar{h}}}^{M}](\mbox{\boldmath
$\rho$}_{0}=0;\mbox{\boldmath $k$}_{1},\mbox{\boldmath
$k$}_{2}) &=&\int d\tilde{\mu}~e^{i\mbox{\boldmath
$k$}_{1}\cdot \mbox{\boldmath $\rho$}_{1}+i\mbox{\boldmath
$k$}_{2}\cdot \mbox{\boldmath $\rho$}_{2}}E_{h{\bar{h}}}^{M}(0;%
\mbox{\boldmath $\rho$}_{1},\mbox{\boldmath $\rho$}_{2})\,,
\end{eqnarray}
where $d\mu =d^{2}\mbox{\boldmath $\rho$}_{0}d^{2}\mbox{\boldmath $\rho$}%
_{1}d^{2}\mbox{\boldmath $\rho$}_{2}$, $d\tilde{\mu}=d^{2}%
\mbox{\boldmath
$\rho$}_{1}d^{2}\mbox{\boldmath $\rho$}_{2}$ and $\mathcal{F}$, $\tilde{%
\mathcal{F}}$ are the Fourier transform operators in three and two variables
respectively. In coordinate space the action of the generators on the
eigenfunctions $E_{h{\bar{h}}}^{M}(\mbox{\boldmath $\rho$}_{10},%
\mbox{\boldmath $\rho$}_{20})$ is:
\begin{eqnarray}
M^{+}E_{h{\bar{h}}}^{M}(\mbox{\boldmath $\rho$}_{0};\mbox{\boldmath $\rho$}%
_{1},\mbox{\boldmath $\rho$}_{2}) &=&-\partial _{0}E_{h{\bar{h}}}^{M}(%
\mbox{\boldmath $\rho$}_{0};\mbox{\boldmath $\rho$}_{1},%
\mbox{\boldmath
$\rho$}_{2}) \,,  \label{MmEM} \\
M^{3}E_{h{\bar{h}}}^{M}(\mbox{\boldmath $\rho$}_{0};\mbox{\boldmath $\rho$}%
_{1},\mbox{\boldmath $\rho$}_{2}) &=&(-h-\rho _{0}\partial _{0})E_{h{\bar{h}}%
}^{M}(\mbox{\boldmath $\rho$}_{0};\mbox{\boldmath $\rho$}_{1},%
\mbox{\boldmath $\rho$}_{2}) \,, \\
M^{-}E_{h{\bar{h}}}^{M}(\mbox{\boldmath $\rho$}_{0};\mbox{\boldmath $\rho$}%
_{1},\mbox{\boldmath $\rho$}_{2}) &=&(2h\rho _{0}+\rho _{0}^{2}\partial
_{0})E_{h{\bar{h}}}^{M}(\mbox{\boldmath $\rho$}_{0};\mbox{\boldmath $\rho$}%
_{1},\mbox{\boldmath $\rho$}_{2})\,.
\end{eqnarray}
Note, that the functions $E_{h{\bar{h}}}^{M}(\mbox{\boldmath $\rho$}_{0};%
\mbox{\boldmath $\rho$}_{1},\mbox{\boldmath $\rho$}_{2})$ can be considered
as the Clebsch-Gordon coefficients in the expansion of the product of the
Reggeon wave functions $\varphi (\mbox{\boldmath $\rho$}_{r})$ in the sum of
the irreducible representations

\begin{equation}
\varphi (\mbox{\boldmath $\rho$}_{1})\,\varphi (\mbox{\boldmath $\rho$}%
_{2})=\int_{-\infty }^{\infty }d\nu \sum_{n=-\infty }^{\infty }a_{\nu n}\int
d^{2}\rho _{0}E_{h{\bar{h}}}^{M}(\mbox{\boldmath $\rho$}_{0};%
\mbox{\boldmath
$\rho$}_{1},\mbox{\boldmath $\rho$}_{2})\,O_{h{\bar{h}}}(%
\mbox{\boldmath
$\rho$}_{0})\,,
\end{equation}
where the conformal weights for the principal series of unitary
representations are

\begin{equation}
h=\frac{1}{2}+i\nu +\frac{n}{2}\,,\,\,{\bar{h}=}\frac{1}{2}+i\nu -\frac{n}{2}
\end{equation}
and the coefficients $a_{\nu n}$ are fixed with the use of the normalization
conditions for the wave functions $\varphi (\mbox{\boldmath $\rho$}_{r})$
and $O_{h{\bar{h}}}(\mbox{\boldmath $\rho$}_{0})$. In the irreducible
representations $O_{h{\bar{h}}}(\mbox{\boldmath $\rho$}_{0})$ the M\"{o}bius
group generators are given below

\begin{eqnarray}
M^{+}O_{h{\bar{h}}}(\mbox{\boldmath $\rho$}_{0}) &=&\partial _{0}O_{h{\bar{h}%
}}(\mbox{\boldmath $\rho$}_{0})\,,\,\, \\
M^{3}O_{h{\bar{h}}}(\mbox{\boldmath $\rho$}_{0}) &=&(1-h+\rho _{0}\partial
_{0})O_{h{\bar{h}}}(\mbox{\boldmath $\rho$}_{0})\,\,, \\
M^{-}O_{h{\bar{h}}}(\mbox{\boldmath $\rho$}_{0}) &=&(-2(1-h)\rho _{0}-\rho
_{0}^{2}\partial _{0})O_{h{\bar{h}}}(\mbox{\boldmath $\rho$}_{0})\,.
\end{eqnarray}
The generators of the M\"{o}bius group can be translated into the momentum
space, using the relation $\mbox{\boldmath $\rho$} \cdot \mbox{\boldmath $k$}%
=(\rho ^{\ast }k+\rho k^{\ast })/2$:
\begin{eqnarray}
M^{+}{\tilde{E}}_{h{\bar{h}}}^{M}(\mbox{\boldmath $\rho$}_{0};%
\mbox{\boldmath $k$}_{1},\mbox{\boldmath $k$}_{2}) &=&e^{i(%
\mbox{\boldmath
$k$}_{1}+\mbox{\boldmath $k$}_{2})\cdot \mbox{\boldmath
$\rho$}_{0}}{\tilde{M}}^{+}(\mbox{\boldmath $k$})\,{\tilde{E}}_{h{\bar{h}}%
}^{M}(\mbox{\boldmath $k$}_{1},\mbox{\boldmath $k$}_{2})\,, \\
M^{3}{\tilde{E}}_{h{\bar{h}}}^{M}(\mbox{\boldmath $\rho$}_{0};%
\mbox{\boldmath $k$}_{1},\mbox{\boldmath $k$}_{2}) &=&e^{i(%
\mbox{\boldmath
$k$}_{1}+\mbox{\boldmath $k$}_{2})\cdot \mbox{\boldmath
$\rho$}_{0}}\left( {\tilde{M}}^{3}(\mbox{\boldmath $k$})+\rho _{0}{\tilde{M}}%
^{+}(\mbox{\boldmath $k$})\right) {\tilde{E}}_{h{\bar{h}}}^{M}(%
\mbox{\boldmath $k$}_{1},\mbox{\boldmath $k$}_{2}) \,, \\
M^{-}{\tilde{E}}_{h{\bar{h}}}^{M}(\mbox{\boldmath $\rho$}_{0};%
\mbox{\boldmath $k$}_{1},\mbox{\boldmath $k$}_{2}) &=&e^{i(%
\mbox{\boldmath
$k$}_{1}+\mbox{\boldmath $k$}_{2})\cdot \mbox{\boldmath
$\rho$}_{0}}\,\left( {\tilde{M}}^{-}(\mbox{\boldmath $k$})-2\rho _{0}{\tilde{%
M}}^{3}(\mbox{\boldmath $k$})-\rho _{0}^{2}{\tilde{M}}^{+}(%
\mbox{\boldmath
$k$})\right) {\tilde{E}}_{h{\bar{h}}}^{M}(\mbox{\boldmath $k$}_{1},%
\mbox{\boldmath $k$}_{2})\,,
\end{eqnarray}
where
\begin{eqnarray}
{\tilde{M}}^{+}(\mbox{\boldmath $k$}) &=&-\frac{i}{2}(k_{1}^{\ast
}+k_{2}^{\ast }) \,, \\
{\tilde{M}}^{3}(\mbox{\boldmath $k$}) &=&-(\partial _{k_{1}^{\ast
}}k_{1}^{\ast }+\partial _{k_{2}^{\ast }}k_{2}^{\ast })\,, \\
{\tilde{M}}^{-}(\mbox{\boldmath $k$}) &=&-2i(\partial _{k_{1}^{\ast
}}^{2}k_{1}^{\ast }+\partial _{k_{2}^{\ast }}^{2}k_{2}^{\ast })\,.
\end{eqnarray}

One can calculate also the action of some bilinear combinations of the
generators in the momentum representation

\[
M^{+}M^{-}{\tilde{E}}_{h{\bar{h}}}^{M}(\mbox{\boldmath $\rho$}_{0};%
\mbox{\boldmath $k$}_{1},\mbox{\boldmath $k$}_{2})=e^{i(\mbox{\boldmath $k$}%
_{1}+\mbox{\boldmath $k$}_{2})\cdot \mbox{\boldmath
$\rho$}_{0}}{\tilde{M}}^{+}(\mbox{\boldmath $k$})\left( {\tilde{M}}^{-}(%
\mbox{\boldmath $k$})-2\rho _{0}{\tilde{M}}^{3}(\mbox{\boldmath $k$})- \rho
_{0}^{2}\tilde{M}^{+}(\mbox{\boldmath $k$})\right) {\tilde{E}}_{h{\bar{h}}%
}^{M}(\mbox{\boldmath $k$}_{1},\mbox{\boldmath
$k$}_{2})\,,
\]
\[
\left( M^{3}(M^{3}+1)+M^{+}M^{-}\right) {\tilde{E}}_{h{\bar{h}}}^{M}(%
\mbox{\boldmath $\rho$}_{0};\mbox{\boldmath
$k$}_{1},\mbox{\boldmath $k$}_{2})=e^{i(\mbox{\boldmath $k$}_{1}+%
\mbox{\boldmath $k$}_{2})\cdot \mbox{\boldmath
$\rho$}_{0}}{\tilde{M}}^{2}\,{\tilde{E}}_{h{\bar{h}}}^{M}(%
\mbox{\boldmath
$k$}_{1},\mbox{\boldmath $k$}_{2})\,,
\]
where the Casimir operator in the ''short'' momentum representation is
\begin{equation}
{\tilde{M}}^{2}={\tilde{M}}^{3}(\mbox{\boldmath $k$})\left( {\tilde{M}}^{3}(%
\mbox{\boldmath $k$})+1\right) +{\tilde{M}}^{+}(\mbox{\boldmath $k$})\,{%
\tilde{M}}^{-}(\mbox{\boldmath $k$})=-(\partial _{k_{1}^{\ast }}-\partial
_{k_{2}^{\ast }})^{2}k_{1}^{\ast }k_{2}^{\ast }\,.
\end{equation}
Note that in our notations ${\tilde{M}}^{3}$ is a generator while ${\tilde{M}%
}^{2}$ is the Casimir operator. The algebra of these generators in momentum
space is the same as in coordinate space:
\[
\lbrack {\tilde{M}}_{r}^{+},{\tilde{M}}_{r}^{3}]={\tilde{M}}_{r}^{+}\,,\quad
\lbrack {\tilde{M}}_{r}^{+},{\tilde{M}}_{r}^{-}]=-2{\tilde{M}}%
_{r}^{3}\,,\quad \lbrack {\tilde{M}}_{r}^{-},{\tilde{M}}_{r}^{3}]=-{\tilde{M}%
}_{r}^{-}\,.
\]

While the action of the generator ${\tilde{M}}^{+}$ does not need special
comments, it is interesting to look at others. One finds
\begin{eqnarray}
{\tilde{M}}^{3}{\tilde{E}}_{h}^{M}(\mbox{\boldmath $k$}_{1},%
\mbox{\boldmath
$k$}_{2}) &=&-h\,{\tilde{E}}_{h}^{M}(\mbox{\boldmath $k$}_{1},%
\mbox{\boldmath $k$}_{2}) \,,  \label{momM3M} \\
{\tilde{M}}^{-}{\tilde{E}}_{h}^{M}(\mbox{\boldmath $k$}_{1},%
\mbox{\boldmath
$k$}_{2}) &=&0\,.  \label{momMminusM}
\end{eqnarray}
The last relation follows from the action of the Casimir operator
\begin{equation}
{\tilde{M}}^{2}{\tilde{E}}_{h}^{M}=h(h-1){\tilde{E}}_{h}^{M}
\label{casimiraction}
\end{equation}
and from relation (\ref{momM3M}). In particular, given that
\begin{equation}
{\tilde{M}}^{2}=\frac{1}{2}\left[ {\tilde{M}}^{+}{\tilde{M}}^{-}+{\tilde{M}}%
^{-}{\tilde{M}}^{+}\right] +\left( {\tilde{M}}^{3}\right) ^{2}={\tilde{M}}%
^{+}{\tilde{M}}^{-}+{\tilde{M}}^{3}\left( {\tilde{M}}^{3}+1\right) \,,
\label{casimir}
\end{equation}
where the commutation relation $[{\tilde{M}}^{+},{\tilde{M}}^{-}]=-2{\tilde{M%
}}^{3}$ has been used, one can see that indeed the relation
(\ref{momMminusM} ) is satisfied.

These relations mean that in the M-space in momentum representation the
M\"obius algebra is projected to
\begin{eqnarray}
[{\tilde M}^+,{\tilde M}^3] {\tilde E}^M ={\tilde M}^+ {\tilde E}^M\, ,
\quad {\tilde M}^-{\tilde M}^+ {\tilde E}^M =2 {\tilde M}^3 {\tilde E}^M \,
, \quad 0\, {\tilde E}^M=0\, {\tilde E}^M\,.
\end{eqnarray}

Let us now move to the $AF$ space. We start, in coordinate space, from the
definition
\begin{equation}
E^A_h=E^M_h-E^\delta_h\,,
\end{equation}
where, according to eq. (\ref{mapping_on_E}),
\begin{equation}
E^\delta_h(\brho_{10},\brho_{20})=2^{1-h-\bar{h}} P
E^M_h(\brho_{10},\brho_{20})= \left(\frac{1}{\rho_{20}}\right)^h \left(\frac{1%
}{\rho_{20}^{*}}\right)^{\bar{h}} + \left(\frac{1}{-\rho_{10}}\right)^h
\left(\frac{1}{(-\rho_{10})^*}\right)^{\bar{h}}
\end{equation}
and $P$ is a projector ($P^2=P$) defined by its action
\begin{equation}
P f(x_1,x_2) = \frac{1}{2}\left( f(x_1,-x_1)+f(-x_2,x_2)\right) \,.
\end{equation}
In order to compute the action of the M\"obius generators on the analytic
part, $E^A_h$, of the conformal eigenfunction we first examine the singular
piece, $E^\delta_h$.

First of all let us observe that the Casimir operator acting on $%
E_{h}^{\delta }$ gives zero, as it
is trivially checked in the coordinate
representation. This means that $h$ in such a case can be obtained with the
action of a function of the Casimir (solving the eq. (\ref{casimiraction})
for $h$) only before applying the projector $P$. After the application of $P$
one needs to define a new operator to extract the scaling parameter $h$. It
is related to the fact, that $E_{h}^{A}$ is not an eigenfunction of the
usual Casimir operator. We should construct a new Casimir operator ${\tilde{M%
}}_{AF}^{2}$ for this representation.

In coordinate representation we can deduce the following
relations for generators
\begin{eqnarray} M^+ E^\delta_h(\brho_{10},\brho_{20})
= -\partial_0 E^\delta_h(\brho_{10},\brho_{20})\,, \\ M^3
E^\delta_h(\brho_{10},\brho_{20}) = (-h - \rho_0 \partial_0)
E^\delta_h(\brho_{10},\brho_{20})\,,  \label{M3deltacoord}
\end{eqnarray}
which coincides with
their
action for $E^M_h$ while the action of $M^-$ is
different. According to this fact the action of M\"obius generators on $%
E_h^A $ can be written as
\begin{eqnarray}
M^+ E^A_h(\brho_{10},\brho_{20}) = -\partial_0 E^A_h(\brho_{10},\brho_{20})\,, \\
M^3 E^A_h(\brho_{10},\brho_{20}) = (-h - \rho_0 \partial_0)
E^A_h(\brho_{10},\brho_{20})
\end{eqnarray}
and
\begin{eqnarray}
M^- E^A_h(\brho_{10},\brho_{20}) =(\rho_0^2 \partial_0 +2 h \rho_0)
E^A_h(\brho_{10},\brho_{20}) - \left[ h \rho_{10} E^M_h(\brho_{10},-\brho_{10})
+ h \rho_{20} E^M_h(-\brho_{20},\brho_{20}) \right] 2^{-h-{\bar{h}}}\,.
\end{eqnarray}
Therefore
only the last generator
$M^-$, needs to be redefined in the $AF$ space.

Let us consider now again the momentum space for these functions. First we
analyze the action of the ${\tilde M}^3$ generator on ${\tilde E}^A_h(%
\mbox{\boldmath $k$}_1,\mbox{\boldmath $k$}_2)$, which we have written in
terms of hypergeometric functions. One can immediately see that
\begin{equation}
{\tilde M}^3 {\tilde E}^A_h(\mbox{\boldmath $k$}_1,\mbox{\boldmath $k$}_2) =
-h \, {\tilde E}^A_h(\mbox{\boldmath $k$}_1,\mbox{\boldmath $k$}_2)
\label{act1} \\
\end{equation}
for any $\mbox{\boldmath $k$}_1+\mbox{\boldmath $k$}_2$. This means again
that we are not forced to use a non local operator $\hat{h}$ (solving an
equation with the corresponding Casimir ${\tilde M}^2_{AF}$) to extract the
dimension $h$ of an eigenstate, as we had to do in the coordinate
representation.

The $E_h^\delta$ function is instead very peculiar since we know that the
Casimir gives zero. This fact is also verified directly in momentum space,
due to the deltas present in ${\tilde E}_h^\delta$. Later we will show it
explicitely. Instead we note that the action of the ${\tilde M}^3$ operator
is good. Indeed, from the explicit form given in Eq. (\ref{pom_delta}), we
obtain
\begin{eqnarray}
{\tilde M}^3 E_h^\delta(\mbox{\boldmath $k$}_1,\mbox{\boldmath $k$}_2)=-h \,
E_h^\delta(\mbox{\boldmath $k$}_1,\mbox{\boldmath $k$}_2)\,.
\end{eqnarray}
This is compatible with the fact that the ${\tilde M}^3$ operator has the
same action on the ${\tilde E}^M_h(\mbox{\boldmath $k$}_1,%
\mbox{\boldmath
$k$}_2)$ and ${\tilde E}^A_h(\mbox{\boldmath $k$}_1,\mbox{\boldmath $k$}_2)$
functions.

The fact that we may choose to extract the dimension $h$ in momentum space
using the ${\tilde M}^3$ operator may be useful to simplify the construction
of the mappings $\Phi$ and $\Phi^{-1}$. On defining
\begin{equation}
F=2^{1-\hat{h}-\hat{{\bar{h}}}}
\end{equation}
one may write the mapping introduced in eq. (\ref{mapping_on_E}) as
\begin{eqnarray}
\Phi^{-1}:M \to AF , \quad \Phi^{-1}= 1 -F^\delta P =1-P F^{M}\,, \\
\Phi:AF \to M , \quad \Phi= 1 +\frac{F^\delta}{1-F^\delta} P=1+P \frac{F^{AF}%
}{1-F^{AF}}\,,
\end{eqnarray}
where in general $P F P=F P$ and which imply $\Phi \Phi^{-1}=
\Phi^{-1}\Phi=1 $ and we have denoted $F^\delta=F(\hat{h}({\tilde M}^3))$, $%
F^{M}=F(\hat{h}({\tilde M}^2))$ and similarly $F^{AF}=F(\hat{h}({\tilde M}%
^2_{AF}))$. Clearly in the relation for the mappings one can use different
forms for the operator $F$ depending on which space it acts.

In momentum space $\hat{h}=-{\tilde M}^3$ then one can
write $[P,F]=0$ (with
some abuse of notation) and things are further simplified.

Proceeding along this line one can give a deformed representation ${\tilde M}%
_{AF}^-$ which lives on the AF-space. In order to do this, let us first
study
\begin{equation}
{\tilde M}^- {\tilde E}^A_h = {\tilde M}^- ({\tilde E}^M_h -{\tilde E}%
^\delta_h) = - {\tilde M}^- {\tilde E}^\delta_h  \label{MminusA}
\end{equation}
and
\begin{eqnarray}
{\tilde M}^- {\tilde E}^\delta_h&=&-2i
(\partial_{k_1^*}^2k_1^*+\partial_{k_2^*}^2k_2^*) \left[\delta^{(2)}(%
\mbox{\boldmath $k$}_1)+(-1)^n \delta^{(2)}(\mbox{\boldmath $k$}_2) \right]
c_h (k_1+k_2)^{{\bar{h}}-1} (k_1^*+k_2^*)^{h-1}  \nonumber \\
&=&2i h (h-1) \frac{1}{k_1^*+k_2^*} {\tilde E}^\delta_h \,=\, -h(h-1) \left(
{\tilde M}^+ \right)^{-1} {\tilde E}^\delta_h  \nonumber \\
&=&-\left( {\tilde M}^+ \right)^{-1} h(h-1) \frac{F}{1-F} P {\tilde E}_h^A
\,.  \label{rel1}
\end{eqnarray}
Note that ${\tilde M}^+$ has the same form on any of the spaces of functions
we consider. The resulting action of ${\tilde M}^-$ on ${\tilde E}^A_h$ has
to be studied in order to see if it is possible to define a new
representation for ${\tilde M}^-$ on AF-space which satisfies the conformal
algebra. But before let us note that from the previous relation we
have
\begin{eqnarray}
{\tilde M}^+ {\tilde M}^- {\tilde E}^\delta_h = -h(h-1) {\tilde E}^\delta_h
\,,  \nonumber \\
{\tilde M}^- {\tilde M}^+ {\tilde E}^\delta_h = -h(h+1) {\tilde E}^\delta_h
\,,  \label{reldelta}
\end{eqnarray}
so that we reobtain, recalling eq. (\ref{casimir}),
\begin{equation}
{\tilde M}^2 {\tilde E}^\delta_h = 0\,.
\end{equation}
In general we have defined the M\"obius generators on the AF space (spanned
by the Basis functions ${\tilde E}^A_h$) using the isometry with the
M-space:
\begin{equation}
{\tilde M}_{AF}=\Phi^{-1} {\tilde M} \Phi\, .
\end{equation}
Let us look at each of the three generators. From the previous discussion,
also in the coordinate representation, we observe that
\begin{equation}
{\tilde M}_{AF}^+={\tilde M}^+ \quad, \quad {\tilde M}_{AF}^3={\tilde M}^3
\,,
\end{equation}
which means that in momentum space the actions of these two operators is the
same in both M-space and AF-space.

We therefore define for the AF-space a new generator in momentum
representation such that
\begin{equation}
{\tilde M}_{AF}^-{\tilde E}^A_h =0\,,
\end{equation}
which puts into relation, as before, the action of the Casimir
operator such that
\begin{equation} {\tilde M}_{AF}^2 {\tilde
E}^A_h =h(h-1) {\tilde E}^A_h \,,
\end{equation}
together with
the piece of the M\"obius algebra $[{\tilde M}_{AF}^+,{\tilde M%
}_{AF}^-]=-2{\tilde M}_{AF}^3$ and the action of ${\tilde M}_{AF}^3$.
According to the previous result (see eqs. (\ref{MminusA}) and (\ref{rel1}%
)), by construction the choice is
\begin{equation}
{\tilde M}_{AF}^-={\tilde M}^- -\left( {\tilde M}^+ \right)^{-1} h(h-1)
\frac{F}{1-F} P= {\tilde M}^- - \left( {\tilde M}^+ \right)^{-1} P \, G({%
\tilde M}_{AF}^2) \,,
\end{equation}
where, in the last form, G is a non local operator such that
\begin{equation}
G {\tilde E}_h^A=h(h-1)\frac{F(h)}{1-F(h)} {\tilde E}_h^A\,.
\end{equation}

Let us finally verify that, with this new representation of ${\tilde M}%
_{AF}^-$, the algebra of the M\"obius generators is correct. It is easy to
see that
\begin{equation}
[{\tilde M}_{AF}^-, {\tilde M}_{AF}^3]=-{\tilde M}_{AF}^- \,,
\end{equation}
which is a relation that is projected to zero when applied to the function ${%
\tilde E}^A_h$. In order to check that the M\"obius algebra is closed we
need to examine the relation
\begin{equation}
[{\tilde M}_{AF}^-, {\tilde M}_{AF}^+]=+2{\tilde M}_{AF}^3
\end{equation}
understood to act on ${\tilde E}^A_h$. One then has to verify that
\begin{equation}
{\tilde M}_{AF}^- {\tilde M}_{AF}^+{\tilde E}^A_h=-2h{\tilde E}^A_h \,.
\label{v1}
\end{equation}
Using eq. (\ref{reldelta}) we derive also
\begin{equation}
{\tilde M}^- {\tilde M}^+ {\tilde E}_h^A = \left( {\tilde M}^+ {\tilde M}^-
+2{\tilde M}^3\right)\left({\tilde E}_h^M-{\tilde E}_h^\delta \right)= -2h {%
\tilde E}_h^A +h(h-1) {\tilde E}^\delta_h  \label{v2}
\end{equation}
and making the difference between eqs. (\ref{v1}) and (\ref{v2}) we
therefore obtain
\begin{equation}
\left( {\tilde M}_{AF}^- - {\tilde M}^- \right) {\tilde M}^+ {\tilde E}_h^A
=-h(h-1) {\tilde E}^\delta_h \,.
\end{equation}
If ${\tilde M}^+$ commutes with $( {\tilde M}_{AF}^- - {\tilde M}^-)$ this
is equivalent to
\begin{equation}
\left( {\tilde M}_{AF}^- - {\tilde M}^- \right) {\tilde E}_h^A = -h(h-1)
\left({\tilde M}^+ \right)^{-1} {\tilde E}^\delta_h = - \left({\tilde M}^+
\right)^{-1} P \, G({\tilde M}_{AF}^2) {\tilde E}_h^A \,,  \label{v3}
\end{equation}
which coincides with the starting definition of the generator. The
commutation relation is fullfilled since the operator ${\tilde M}^+$ is just
a multiplicative operator associated to the total momentum and therefore
commutes with the action of the projector $P$ and moreover it commutes with
the function of the Casimir $G({\tilde M}_{AF}^2)$.

Finally let us look at the relation between the two Casimir operators in
M-space and AF-space: using ${\tilde{M}}_{AF}^{2}=\frac{1}{2}\left[ {\tilde{M%
}}_{AF}^{+}{\tilde{M}}_{AF}^{-}+{\tilde{M}}_{AF}^{-}{\tilde{M}}_{AF}^{+}%
\right] +\left( {\tilde{M}}_{AF}^{3}\right) ^{2}$ we find from the previous
relations
\begin{equation}
{\tilde{M}}_{AF}^{2}+P\,G({\tilde{M}}_{AF}^{2})={\tilde{M}}^{2}\,.
\end{equation}
This last relation, applied to ${\tilde{E}}_{h}^{A}$, gives
\begin{equation}
h(h-1){\tilde{E}}_{h}^{A}+h(h-1){\tilde{E}}_{h}^{\delta }={\tilde{M}}^{2}{%
\tilde{E}}_{h}^{A}\,,
\end{equation}
which indeed is compatible with ${\tilde{M}}^{2}{\tilde{E}}_{h}^{M}=h(h-1){%
\tilde{E}}_{h}^{M}$.
%%%%%%%%%%%%%%%%%%%%%%%%%%%%%%%%%%%%%%%%%%%%%%%%%%%%%%%%%%%%%%%%%%%%%%%%%%

\section{Conclusions}

Among the properties of the LL BFKL kernel the bootstrap relation
connected to the gluon
reggeization is a fundamental consistency condition. This bootstrap property
leads to an important feature of the BFKL kernel also in the color singlet
state: this can be demonstrated most easily in momentum space where the BFKL
kernel is meromorphic. Many previous investigations, exploiting the
conformal symmetry of the kernel, have been carried out in the space of
M\"obius functions ($M$). In this space of functions, however, the bootstrap
relation for the BFKL kernel is not satisfied.

In this paper we have defined a modified space of functions ($AF$), in which
the bootstrap property is valid. In particular, we have derived a
spectral representation of the kernel which is also useful in evaluating the
coupling of the BFKL Green's function to colored impact factors, and we have
verified explicitly that the bootstrap relation is fulfilled.
We also have
derived the corresponding represention of the M\"obius algebra; in
order to act in this modified space of functions, one of the
M\"obius generators has to be deformed.

Our discussion has been limited to the the case of two-gluon Green's
functions. Since in both spaces, $M$ and $AF$, the eigenvalues of the
Casimir operator, ${\tilde{M}}^{2}$ and ${\tilde{M}}_{AF}^{2}$ , resp., are
the same, and since the hamiltonian for the pairwise interaction of two
reggeized gluons can be expressed in terms of this Casimir operator, one can
expect that, for states consisting of more than two gluons all remarkable
properties of the multi-colour BFKL dynamics derived in the M\"{o}bius
picture can be generalized to the AF picture. This includes, in particular,
the holomorphic separability and integrability \cite{int}. We
are going to return to these interesting problems in future
publications.

%%%%%%%%%%%%%%%%%%%%%%%%%%%%%%%%%%%%%%%%%%%%%%%%%%%%%%%%%%%%%%%%%%%%%%%

\section*{Acknowledgements}

Part of this work has been done while one of us (L.N.L) has been visiting
the II.Institut f.Theoretische Physik, University Hamburg, and the
University of Bologna, Bologna. Two of us (J.B. and G.P.Vacca) gratefully
acknowledge the hospitality of the Nuclear Physics Institute in Gatchina,
St.Petersburg.
%%%%%%%%%%%%%%%%%%%%%%%%%%%%%%%%%%%%%%%%%%%%%%%%%%%%%%%%%%%%%%%%%%%%%%%

\section*{Appendix A}

The calculation
of the integral in eq. (\ref{integral}) is carried on noting
that one can find a third order differential operator in $z$ and
another in $ z^*$, related to the ${}_3 F_2$ functions, which,
when applied to the integrand, gives a total derivative in an
integration variable so that, on applying the Stokes theorem, one
finds that $I_\lambda(z)$ satisfies the related differential
equation in both the holomorphic and antiholomorphic sectors.
This allows to write it as a sum of products of the linearly
independent solutions of the two differential equations and on imposing the
single valueness and the correct normalization (at some convenient point)
the full expression is found \cite{geronimo}. In particular this has the
following, so called, conformal block structure
\begin{equation}
I_\lambda(z)= \sum_{i=0}^2 \lambda_i \, u_i(z) \bar{u}_i(z^*) \, ,
\end{equation}
where $u_i$ and $\bar{u}_i$ are three independent solutions of the
generalized hypergeometric differential equations in $z$ and $z^*$. The
coefficients $\lambda_i$ depend through $\Gamma$ functions on the conformal
weights. The general form of $I_{\lambda}(z)$ is pretty complicated but it
simplifies considerably for $z=1$ using the relations~\cite{Bateman}
\begin{eqnarray}
{}_3F_2 \left({
\begin{array}{c}
a,b,c\\
\frac{a+b+1}{2},2 c
\end{array}
}\bigg|1\right) &=&
\pi^{1/2} \frac{\Gamma \left(c+\frac{1}{2}\right) \Gamma \left(\frac{1+a+b}{2%
}\right) \Gamma \left(c+\frac{1-a-b}{2}\right) } {\Gamma \left(\frac{1+a}{2}%
\right) \Gamma \left(\frac{1+b}{2}\right) \Gamma \left(c+\frac{1-a}{2}%
\right) \Gamma \left(c+\frac{1-b}{2}\right) } \, ,  \nonumber \\
{}_3F_2 \left({
\begin{array}{c}
a,1-a,c\\
f,2 c+1-f
\end{array}
}\bigg|1\right) &=& \pi \frac{%
\Gamma \left(f\right) \Gamma \left(2c+1-f\right) 2^{1-2c}} {\Gamma \left(%
\frac{a+f}{2}\right) \Gamma \left(\frac{1-a+f}{2}\right) \Gamma \left(c+%
\frac{1+a-f}{2}\right) \Gamma \left(1+c-\frac{a+f}{2}\right) } \, .
\end{eqnarray}
On using the above relation, after algebraic simplifications, we find
\begin{equation}
I_{\lambda}(1)=I^{(0)}+I^{(1)}+I^{(2)}  \label{resultI}
\end{equation}
with
\begin{eqnarray}
I^{(0)}&=& \frac{2^{-3+4\lambda} \Gamma \left(1-\frac{h}{2}\right) \Gamma
\left(1-\frac{\bar{h}}{2}\right) \Gamma^2 \left(1-\lambda \right) \Gamma^2
\left(-\lambda \right) \Gamma \left(\frac{1-h}{2}+\lambda\right) \Gamma
\left(\frac{1-\bar{h}}{2}+\lambda\right) } {\Gamma \left(\frac{3-h}{2}%
\right) \Gamma \left(\frac{3-\bar{h}}{2}\right) \Gamma \left(1-\frac{h}{2}%
-\lambda\right) \Gamma \left(1-\frac{\bar{h}}{2}-\lambda\right) } \times
\nonumber \\
&& \times \frac{ \sin{\pi(\bar{h}-2\lambda)}\sin{\pi \lambda} \tan{\pi \bar{h%
}} } {\sin{\pi(\bar{h}-\lambda)}} \, ,  \nonumber \\
I^{(1)}&=&- \frac{2^{-3+4\lambda} \Gamma \left(\frac{h-1}{2}\right) \Gamma
\left(\frac{\bar{h}-1}{2}\right) \Gamma^2 \left(1-\lambda \right) \Gamma^2
\left(-\lambda \right) \Gamma \left(\frac{h}{2}+\lambda\right) \Gamma \left(%
\frac{\bar{h}}{2}+\lambda\right) } {\Gamma \left(\frac{h}{2}\right) \Gamma
\left(\frac{\bar{h}}{2}\right) \Gamma \left(\frac{1+h}{2}-\lambda\right)
\Gamma \left(\frac{1+\bar{h}}{2}-\lambda\right) } \times  \nonumber \\
&& \times \frac{ \sin{\pi(\bar{h}+2\lambda)}\sin{\pi \lambda} \tan{\pi \bar{h%
}} } {\sin{\pi(\bar{h}+\lambda)}} \, ,  \nonumber \\
I^{(2)}&=& \frac{2^{4\lambda}\pi^{6} \, \Gamma^2 \left(1-\lambda \right) } {%
\Gamma \left(\frac{3-h}{2}\right) \Gamma \left(\frac{h}{2}\right) \Gamma
\left(\frac{3-\bar{h}}{2}\right) \Gamma \left(\frac{\bar{h}}{2}\right)
\Gamma \left(\frac{1+h}{2}-\lambda\right) \Gamma \left(\frac{1+\bar{h}}{2}%
-\lambda\right) } \times  \nonumber \\
&& \times \frac{1}{ \Gamma \left(1-\frac{h}{2}-\lambda\right) \Gamma \left(1-%
\frac{\bar{h}}{2}-\lambda\right) \Gamma^2 \left(1+\lambda \right) } \frac{1}{%
\sin{\pi(\bar{h}-\lambda)}\sin{\pi(\bar{h}+\lambda)}} \, .
\end{eqnarray}
The expressions of $I^{(0)}$ and $I^{(1)}$ have a simple pole at $\lambda=0$
and opposite residues for this pole, while $I^{(2)}$ is holomorphic at $%
\lambda=0$, therefore the full function in eq. (\ref{resultI}) is not
singular at $\lambda=0$. Using the transformation properties of the Gamma
function, each of
three terms can be reduced to a common multiplicative factor
and expressions involving only trigonometric functions. These terms
are combined
together to give the following compact form for $I_{\lambda}(1)$
\begin{eqnarray}
I_{\lambda}(1)&=& \frac{2}{(1-h)(1-\bar{h})} \, \delta_{n,\mathrm{even}} \,
\times  \nonumber \\
&&\frac{ \Gamma \left(\frac{h}{2}+\lambda\right) \, \Gamma \left(\frac{\bar{h%
}}{2}+\lambda\right) \, \Gamma \left(\frac{1-h}{2}+\lambda\right) \, \Gamma
\left(\frac{1-\bar{h}}{2}+\lambda\right) }{ \Gamma \left(\frac{h}{2}\right)
\, \Gamma \left(\frac{\bar{h}}{2}\right) \, \Gamma \left(\frac{1-h}{2}%
\right) \, \Gamma \left(\frac{1-\bar{h}}{2}\right) } \, \times  \nonumber \\
&&16^{\lambda}\left(1+i^{n} \, \frac{\sin{2 \pi \lambda }}{\sin{\ \pi \left(%
\frac{h+\bar{h}}{2} \right)}}\right) \, \Gamma^2 \left(1-\lambda \right) \,
\Gamma^2 \left(-\lambda \right) \, \sin^{2}{\pi \lambda}.
\end{eqnarray}
Now the first few derivatives with respect to $\lambda$ can be easily
calculated.
In particular we obtain, after some manipulations,
\begin{eqnarray}
c_0 \equiv I_{\lambda}(1)\arrowvert_{\lambda=0}&=& \frac{2\pi^2}{(1-h)(1-%
\bar{h})} \, \, \delta_{n,\mathrm{even}} \, ,  \nonumber \\
I^{\prime}_{\lambda}(1)\arrowvert_{\lambda=0}&=& 2 \, \epsilon_h \, c_0 \, ,
\nonumber \\
I^{\prime\prime}_{\lambda}(1)\arrowvert_{\lambda=0}&=& \bigg[ \left( 2 \,
\epsilon_h \right)^2 - \left(-1\right)^{h-\bar{h}} \, 4\pi^2 \, \csc^2{\frac{%
h+\bar{h}}{2}} +  \nonumber \\
&& \psi^{\prime}\left( \frac{h}{2} \right) + \psi^{\prime}\left( \frac{\bar{h%
}}{2} \right) + \psi^{\prime}\left( \frac{1-h}{2} \right) +
\psi^{\prime}\left( \frac{1-\bar{h}}{2} \right) \bigg] \, c_0 \,.
\label{d01}
\end{eqnarray}
%%%%%%%%%%%%%%%%%%%%%%%%%%%%%%%%%%%%%%%%%%%%%%%%%%%%%%%%%%%%%%%%%%%%%%%%%%

\section*{Appendix B}

We recompute here the double iteration of the BFKL kernel in momentum space
acting
on a constant function~\cite{BaKu} and compare it with
the expressions obtained with the use of the above
spectral representation of the kernel (and therefore of the
Green's function). Let us write before the result of the direct
integration in momentum space:  \begin{eqnarray}
(\bar{K}_{12}^{(1)})^2\otimes 1 &=& \frac{1}{2} \bar{\alpha}_s \, \bar{K}%
_{12}^{(1)} \otimes \log\left( \frac{\mbox{\boldmath $q$}^4}{\mbox{\boldmath
$k$}_1^2 \mbox{\boldmath $k$}_2^2} \right) \, =  \nonumber \\
&=& \left( \frac{1}{2} \bar{\alpha}_s \right)^2 \left[ \log^2\left( \frac{%
\mbox{\boldmath $q$}^2}{\mbox{\boldmath $k$}_1^2} \right) + \log^2\left(
\frac{\mbox{\boldmath $q$}^2}{\mbox{\boldmath $k$}_2^2} \right) \right] \, .
\label{Ksquare1analytic}
\end{eqnarray}
Before giving some details of the computation let us consider what we obtain
from the spectral approach. On using the spectral representation for the
kernel, the relation (\ref{Ksquare1analytic}) can be written as
\begin{equation}
\langle k | \left( \hat{\bar{K}}_{12}^{(1)} \right)^2 | U \rangle = \left(
\bar{\alpha}_s \right)^2 \sum_h \tilde{N}_h \, \langle k | h^{A} \rangle \,
\epsilon_h^2 \, \langle h^{A} | U \rangle \, ,  \label{Ksquare1spectral}
\end{equation}
a relation which we have checked numerically. This check has been also
performed in
the case of the single iteration (bootstrap relation). The
argument of the integration over $\nu$ oscillates with a very slow decay for
fixed conformal spin, but summing over some tens of $n$ gives a good
suppression of the tails and the integral can be computed.

Finally we sketch the derivation of eq. (\ref{Ksquare1analytic}).
There are three kinds of integral involved; they can be calculated with
the dimensional regularization and the
Feynman parameterization, in the case $%
\mbox{\boldmath $k$}_{1,2} \ne 0, \mbox{\boldmath $k$}_1 \ne \mbox{\boldmath
$k$}_2$.\newline
The needed integrals are:
\begin{eqnarray}
\omega (\mbox{\boldmath $k$}_1) &=& \int d^d\mbox{\boldmath $k$}^{\prime}\,
\frac{\mbox{\boldmath $k$}_1^2}{
(\mbox{\boldmath $k$}^{\prime})^{\,2} (%
\mbox{\boldmath $k$}_1-\mbox{\boldmath $k$}^{\prime})^2} \, ,  \nonumber \\
J_1 (\mbox{\boldmath $k$}_1,\mbox{\boldmath $k$}_2) &=& \int d^d%
\mbox{\boldmath $k$}^{\prime}\, \frac{\mbox{\boldmath $k$}_1^2}{%
(\mbox{\boldmath $k$}^{\prime})^{\,2} 
(\mbox{\boldmath $k$}_1-\mbox{\boldmath
$k$}^{\prime})^2} \log \frac{\mbox{\boldmath $q$}^2}{
(\mbox{\boldmath $k$}^{\prime})^{\,2}} \, , \\
J_2 (\mbox{\boldmath $k$}_1,\mbox{\boldmath $k$}_2) &=& \int d^d%
\mbox{\boldmath $k$}^{\prime}\, \frac{\mbox{\boldmath $k$}_1^2}{%
(\mbox{\boldmath $k$}^{\prime})^{\,2} 
(\mbox{\boldmath $k$}_1-\mbox{\boldmath
$k$}^{\prime})^2} \log \frac{\mbox{\boldmath $q$}^2}{(\mbox{\boldmath $q$}-%
\mbox{\boldmath $k$}^{\prime})^{\,2}} \, ,  \nonumber
\end{eqnarray}
where $d=2+2\,\epsilon$, $\mbox{\boldmath $q$} = \mbox{\boldmath $k$}_1+%
\mbox{\boldmath $k$}_2$ and $\omega (\mbox{\boldmath $k$}_1)$ is the gluon
trajectory (\ref{gluontrajectory}) rescaled by $-N_c/2 c$.\newline
The first two integral can be easily calculated exactly, giving:
\begin{eqnarray}
\omega (\mbox{\boldmath $k$}_1) &=& \pi^{1+\epsilon} \frac{%
\Gamma(1-\epsilon)\Gamma^2(\epsilon)} {\Gamma(2 \epsilon)} (\mbox{\boldmath
$k$}_1^2)^\epsilon =  \nonumber \\
&=& \pi^{1+\epsilon} \Gamma(1-\epsilon) (\mbox{\boldmath $q$}^2)^\epsilon %
\left[ \frac{2}{\epsilon} -2\log \frac{\mbox{\boldmath $q$}^2}{%
\mbox{\boldmath $k$}_1^2} + \mathcal{O}(\epsilon) \right] \, ,  \nonumber \\
J_1 (\mbox{\boldmath $k$}_1,\mbox{\boldmath $k$}_2) &=& \pi^{1+\epsilon}
\frac{\Gamma(1-\epsilon)\Gamma^2(\epsilon)} {\Gamma(2 \epsilon)} \left[ \pi
\cot \pi \epsilon + \log \frac{\mbox{\boldmath $q$}^2}{\mbox{\boldmath $k$}%
_1^2} + \psi(2 \epsilon) - \psi(1) \right] (\mbox{\boldmath $k$}%
_1^2)^\epsilon =  \nonumber \\
&=& \pi^{1+\epsilon} \Gamma(1-\epsilon) (\mbox{\boldmath $q$}^2)^\epsilon %
\left[ \frac{1}{\epsilon^2} + \frac{1}{\epsilon} \log \frac{\mbox{\boldmath
$q$}^2}{\mbox{\boldmath $k$}_1^2} - \frac{\pi^2}{6} -\frac{3}{2}\log^2 \frac{%
\mbox{\boldmath $q$}^2}{\mbox{\boldmath $k$}_1^2} + \mathcal{O}(\epsilon) %
\right] \, .
\end{eqnarray}
The third integral is calculated to the order
$\mathcal{O}(\epsilon)$:
\begin{eqnarray} J_2 (\mbox{\boldmath
$k$}_1,\mbox{\boldmath $k$}_2) &=& \pi^{1+\epsilon}
\Gamma(1-\epsilon) (\mbox{\boldmath $q$}^2)^\epsilon \left[ \frac{1}{\epsilon%
} \log \frac{\mbox{\boldmath $q$}^2}{\mbox{\boldmath $k$}_2^2} - \frac{1}{2}
\log \frac{\mbox{\boldmath $q$}^2}{\mbox{\boldmath $k$}_2^2} - \log{\frac{%
\mbox{\boldmath $q$}^2}{\mbox{\boldmath $k$}_1^2}} \log{\frac{%
\mbox{\boldmath $q$}^2}{\mbox{\boldmath $k$}_2^2}}+ \mathcal{O}(\epsilon) %
\right] \, .
\end{eqnarray}
Putting together all the pieces to construct the complete action of the
kernel, all the singulatities cancel and the result, which is finite in the
limit $\epsilon \rightarrow 0$, correspond to the expression given in eq. (%
\ref{Ksquare1analytic}).

%%%%%%%%%%%%%%%%%%%%%%%%%%%%%%%%%%%%%%%%%%%%%%%%%%%%%%%%%%%%%%%%%%%%%%%%%%

%References...


\begin{thebibliography}{99}
\bibitem{BFKL}  E. A. Kuraev, L. N. Lipatov and V. S. Fadin,  Sov. \textbf{%
JETP 44} (1976) 443; \newline
\textbf{ibid. 45} (1977) 199;\newline
Ya. Ya. Balitskii and L.N. Lipatov, Sov. J. Nucl. Phys. \textbf{28}, (1978)
822.

%\cite{Bartels:1999yt}

\bibitem{BLV}  J.~Bartels, L.~N.~Lipatov and G.~P.~Vacca,
%``A new odderon solution in perturbative QCD,''
Phys.\ Lett.\ B \textbf{477} (2000) 178, [arXiv:hep-ph/9912423].
%%CITATION = HEP-PH 9912423;%%

\bibitem{vacca}  G.~P.~Vacca,
%``Properties of a family of n reggeized gluon states in multicolour QCD,''
Phys.\ Lett.\ B \textbf{489}, 337 (2000), [arXiv:hep-ph/0007067].
%%CITATION = HEP-PH 0007067;%%

\bibitem{Lipatov86}  L.~N.~Lipatov,
%``The Bare Pomeron In Quantum Chromodynamics,''
Sov.\ Phys.\ JETP \textbf{63} (1986) 904 [Zh.\ Eksp.\ Teor.\ Fiz.\ \textbf{90%
} (1986) 1536]. %%CITATION = SPHJA,63,904;%%

\bibitem{BLV2}  J.~Bartels, L.~N.~Lipatov and G.~P.~Vacca,
%``Interactions of Reggeized gluons in the Moebius representation,''
Nucl.\ Phys.\ B \textbf{706} (2005) 391, [arXiv:hep-ph/0404110].
%%CITATION = HEP-PH 0404110;%%

\bibitem{int}  L.~N.~Lipatov,
%"High-energy asymptotics of multi-colour QCD and two-dimensional
%conformal field theories''
Phys. Lett. \textbf{B309} (1993) 394; {\it High energy
asymptotics of multi-colour QCD and exactly solvable lattice
models}, hep-th/9311037, unpublished.

\bibitem{Lcft}  L.~N.~Lipatov, \textit{Pomeron in quantum chromodynamics},
in  ``Perturbative QCD'', pp. 411-489, ed. A. H. Mueller,  World Scientific,
Singapore, 1989;  Phys. Rep. \textbf{286} (1997) 131.

\bibitem{dual}  L.~N.~Lipatov,
%"Duality symmetry of the reggeon interactions in the multi-colour QCD"
Nucl.Phys. \textbf{B548} (1999) 328.

\bibitem{dipole}  N.~N.~Nikolaev, B.~G.~Zakharov, Phys.Lett. \textbf{B327}
(1994) 149, 157; A.~N.~Mueller, Nucl.Phys. \textbf{B415} (1994) 377.

\bibitem{BBCV}  J.~Bartels, M.~A.~Braun, D.~Colferai and G.~P.~Vacca,
%``Diffractive eta/c photo- and electroproduction with the
%perturbative QCD$
Eur.Phys.J.C \textbf{20} (2001) 323, [arXiv:hep-ph/0102221].
%%CITATION = HEP-PH 0102221;%%

\bibitem{LV1}  H.~J.~De Vega and L.~N.~Lipatov,
%``Interaction of Reggeized gluons in the Baxter-Sklyanin representation,''
Phys.\ Rev.\ D \textbf{64}, 114019 (2001), [arXiv:hep-ph/0107225],
%%CITATION = HEP-PH 0107225;%%
%``Exact resolution of the Baxter equation for reggeized gluon
%interactions ``
Phys. \ Rev. \ D \textbf{66} (2002) 074013, [arXiv:hep-ph/0107225].

\bibitem{DKKM}  S.~E.~Derkachov, G.~P.~Korchemsky, J.~Kotanski and
A.~N.~Manashov,
%``Noncompact Heisenberg spin magnets from high-energy QCD. II:  Quantization
%conditions and energy spectrum,''
Nucl.\ Phys.\ B \textbf{645} (2002) 237 [arXiv:hep-th/0204124].
%%CITATION = HEP-TH 0204124;%%

%\cite{MT}

\bibitem{MT}   A.~H.~Mueller and W.~K.~Tang,
%``High-energy parton-parton elastic scattering in QCD,''
Phys.\ Lett.\ B \textbf{284} (1992) 123. %%CITATION = PHLTA,B284,123;%%

%\cite{BFLLRW}

\bibitem{BFLLRW}   J.~Bartels, J.~R.~Forshaw, H.~Lotter, L.~N.~Lipatov,
M.~G.~Ryskin and M.~Wusthoff,
%``How does the BFKL pomeron couple to quarks?,''
Phys.\ Lett.\ B \textbf{348} (1995) 589  [arXiv:hep-ph/9501204].
%%CITATION = HEP-PH 9501204;%%

\bibitem{geronimo}  J.~S.~Geronimo and H.~Navelet,
%``On certain two dimensional integrals that appear in conformal
%field  theory,''
J.\ Math.\ Phys.\ \textbf{44} (2003) 2293. %%CITATION = MATH-PH 0003019;%%

\bibitem{Bateman}  Bateman Project, A. Erdelyi, Higher Trascendental
Functions,  Vol. I, pag. 189, McGraw-Hill (1953).

\bibitem{BaKu}   I.~Balitsky and E.~Kuchina,
%``Deeply virtual Compton scattering at small x,''
Phys.\ Rev.\ D \textbf{62} (2000) 074004  [arXiv:hep-ph/0002195].
%%CITATION = HEP-PH 0002195;%%
\end{thebibliography}
\end{document}